\documentclass[aps,prd,floatfix,noshowpacs,10pt,showkeys,tikz,nofootinbib,
nototlepage,twocolumn]{revtex4-2}

\usepackage[toc,page]{appendix}

\makeatletter
\newcommand*{\toccontents}{\@starttoc{toc}}
\makeatother

\usepackage[usenames,dvipsnames]{xcolor}

\definecolor{webgreen}{rgb}{0,0.75,0}
\definecolor{Mgreen}{rgb}{0,0.9,0.8}
\definecolor{webred}{rgb}{0.75,0,0}
\definecolor{webblue}{rgb}{0,0,0.75}
\definecolor{darkblue}{rgb}{0,0,0.7}
\definecolor{dunkelgrau}{rgb}{0.8,0.8,0.8}
\definecolor{lgray}{rgb}{0.95,0.95,0.95}
\definecolor{lgreen}{rgb}{0.95,1.00,0.90}
\definecolor{lblue}{rgb}{0.9,0.95,1.00}
\definecolor{lred}{rgb}{1.00,0.90,0.80}
\definecolor{shadecolor}{rgb}{1.00,0.92,0.82}

\DeclareMathAlphabet{\mathpzc}{OT1}{pzc}{m}{it}

\usepackage{graphicx}

\usepackage{natbib}

\usepackage{tikz-cd}
\usepackage{amssymb}
\usepackage{amsmath}

\tikzcdset{labels={font=\everymath\expandafter{\the\everymath\textstyle}}}

\usepackage{amsfonts}
\usepackage{mathrsfs}
\usepackage{bm}
\usepackage{color}

\usepackage[colorlinks=true,linkcolor=darkblue,citecolor=darkblue,urlcolor=darkblue,breaklinks=true]{hyperref}

\def\d{\mathrm{d}}

\def\i{\mathrm{i}}
\def\e{\mathrm{e}}

\usepackage{orcidlink}

\makeatletter
\newcommand{\fmarki}{*}
\newcommand{\fmarkii}{\ensuremath{\dagger}}
\newcommand{\fmarkiii}{\ensuremath{\ddagger}}
\newcommand{\fmarkiv}{\ensuremath{\mathsection}}
\newcommand{\fmarkv}{\ensuremath{\mathparagraph}}
\newcommand{\fmarkvi}{\ensuremath{\|}}
\newcommand{\fmarkvii}{**}
\newcommand{\fmarkviii}{\ensuremath{\dagger\dagger}}
\newcommand{\fmarkix}{\ensuremath{\ddagger\ddagger}}

 \newcommand{\dobigx}[1]{%
\vcenter{\kern.2ex\hbox{\sffamily#1X}\kern.2ex}}

\def\@fnsymbol#1{{\ifcase#1\or \fmarki\or \fmarkii\or \fmarkiii\or \fmarkiv\or \fmarkv\or \fmarkvi\or \fmarkvii\or \fmarkviii\or \fmarkix \else\@ctrerr\fi}}
\makeatother

\renewcommand{\fmarki}{$\bigstar$}
\renewcommand{\fmarkii}{b$_2$}
\renewcommand{\fmarkiii}{c$_3$}
\renewcommand{\fmarkiv}{a$_4$}
\renewcommand{\fmarkv}{x$_5$}
\renewcommand{\fmarkix}{z$_9$}

\begin{document}

\vspace*{1cm}

\title{
Hamiltonian dynamics of classical spins}

\author{Slobodan Rado\v sevi\' c \orcidlink{0000-0002-3211-1392}\; } 
\email{slobodan@df.uns.ac.rs}
\author{Sonja Gombar \orcidlink{0000-0002-5204-6831}}
\author{Milica Rutonjski \orcidlink{0000-0002-1172-9062}\;}
\author{Petar Mali \orcidlink{0000-0002-3879-5051}}
\author{Milan Panti\' c \orcidlink{0000-0003-4291-8672}}
\author{Milica Pavkov-Hrvojevi\' c \orcidlink{0000-0003-4605-2589}}


\affiliation{Department of Physics, Faculty of Sciences, University of Novi Sad, Trg Dositeja
 Obradovi\' ca 4, Novi Sad, Serbia
}

\begin{abstract}

\noindent
We discuss the geometry behind classical Heisenberg model at the level suitable for third or fourth year students who did not have the opportunity to take a course on differential geometry. The arguments 
presented here rely solely on  elementary    algebraic concepts such as vectors, dual vectors and tensors, as well as  Hamiltonian equations and
Poisson brackets in their simplest form. We derive Poisson brackets for classical spins, along with the corresponding equations of motion for classical
Heisenberg model, starting from the geometry of two-sphere, thereby demonstrating the relevance of standard canonical procedure in the case of Heisenberg model.

\end{abstract}


\maketitle  
\tableofcontents

\section{Introduction}

\noindent Before diving into the quantum world, students
familiarize themselves with classical physics. This is certainly not because classical physics is more fundamental, but rather knowledge of classical physics can be helpful in understanding counterintuitive quantum phenomena. Indeed, the so-called process of quantization, both in Hamiltonian and Lagrangian formulations, relies on the corresponding classical physical system as a starting point in constructing quantum theory.
Examples of this strategy are numerous: the harmonic oscillator is studied as a prime example in classical mechanics with the quantum version being of equal importance. Pauli's solution to the quantum mechanical hydrogen atom is the direct analog of Kepler problem in classical mechanics \cite{WeinbergQM,Schiff}. 
Finally, the prerequisite for learning
quantum electrodynamics is the knowledge of Maxwell's classical theory \cite{Ryder, Schiff,WeinbergQTF1}.

An important exception to this rule of studying quantum system after the classical one is fully understood, is the (quantum) Heisenberg model which captures the physics of localized magnetism. In the simplest case of an isotropic ferromagnet with nearest-neighbor interaction, it is defined by the Hamiltonian operator
\begin{equation}
    \hat H = - \frac J2 \sum_{\bm n, \bm \lambda}\hat{ \bm S}(\bm n) \cdot \hat{\bm S}(\bm n + \bm \lambda)  \label{QuantHeisHamDef}
\end{equation}
where $J>0$ denotes the exchange integral, $\bm n$ is the position vector of a site in the lattice, $\{\bm \lambda\}$ is the set of vectors connecting each
lattice site with its nearest neighbors, $\hat{\bm S}(\bm n)$ is the
spin operator corresponding to the lattice site $\bm n$ and components of spin operators satisfy equal-time commutation relation (we are dropping temporal coordinate for clarity)
\begin{equation}
    \Big{[} \hat{S}_{\alpha}(\bm n), \hat S_{\beta}(\bm m) \Big{]}
= \i \epsilon_{\alpha \beta \gamma} \Delta(\bm n - \bm m) \hat S_{\gamma}(\bm n),
\label{SpinCCR}
\end{equation}
where $\Delta(\bm n-\bm m)=1$ only for $\bm n = \bm m$ and $\Delta(\bm n-\bm m) =0$ in all other cases and $\epsilon_{\alpha \beta \gamma}$ denotes the Levi-Civita
symbol, which is completely antisymmetric tensor in three dimensions with $\epsilon_{123}=1$.
In essence, $\{ \hat S_\alpha(\bm n) \}$ operators from different sites commute, while those on the same site satisfy standard $\mathfrak{su}(2)$ commutation relations. The Hilbert space for this model consists of tensor products of
spin states at all sites \cite{Auerbach}.

When introducing Heisenberg model in
texts for undergraduate students, authors usually start from the
exchange interaction and the Dirac Hamiltonian \cite{Kittel,Nolting},
while graduate-level books discuss the Hubbard model and derive
the Heisenberg Hamiltonian as an effective model which describes 
the physics of spin exchange and superexchange \cite{Auerbach,CMFT,Mattis,Yosida}.
This kind of presentation is, of course, completely reasonable from the
point of view of atomic physics. After all, quantum-mechanical properties
of electrons are key to our understanding
of magnetism. However,
it serves an injustice to the
corresponding model of classical spins since spin waves (or magnons),
as emergent degrees of freedom in a quantum model, are
the quantized version of small collective oscillations of classical spins.
Likewise, a direct confrontation with quantum many-body Hamiltonian \eqref{QuantHeisHamDef} may leave
students wondering what happened to the canonical prescription of replacing
Poisson brackets with commutators, or
how  the spin operators suddenly became dynamical degrees of freedom.
It turns out that all these issues
arise because the underlying phase space for a single classical spin fails to
be a Euclidean space -- it is the two-sphere ${\rm S}^2$ and 
 a full understanding of the classical Heisenberg
model requires certain knowledge of tensor analysis on ${\rm S}^2$.
The problem is that most undergraduates do not
take courses on differential geometry. Still, some elements
of differential geometry are used in physics courses. For example,
introductory texts on quantum field theory describe gauge
field strength as curvature \cite{Ryder,Peskin}, while modern introductory
texts to general relativity cover many important geometric concepts
\cite{RyderGR,ZeeGR}.
It is unfortunate  that the classical Heisenberg model does not get the same treatment, giving the fact that students usually possess needed tools
acquired on linear algebra and vector calculus courses and that
the symplectic form and associate Poisson tensor, which we shall be discussing here, 
are no more abstract than the curvature or parallel transport along a curve
on a manifold. As far as standard textbook treatments are concerned, 
the classical Heisenberg model  remains somewhat mysterious of an object
and students are 
missing a nice opportunity of connecting classical and quantum Heisenberg
models in a compelling way.

The paper is not intended as an introduction to symplectic
geometry, 
nor to the theories of classical and quantum Heisenberg
models; standard references cover these topics in depth \cite{Fecko,Frankel,Auerbach}. Instead, it points to frequently omitted geometry of classical
Heisenberg spins and its connection and relevance to the quantum model.
In the second section we review vectors, dual vectors and tensors,
as well as tensor fields, while the symplectic form and
the Poison tensor on $\mathbb{R}^2$ are introduced in Section \ref{Sect3}.
Generalization of these objects to the case when phase space
is two-sphere is given in Section \ref{Sect4}, which also contains
an example of a simple system defined on the phase space ${\rm S^2}$.
Finally, the application of these ideas to the case of classical
Heisenberg model is given in 
Section \ref{ClassHeisSect}
and  an alternative definition of Poisson brackets in terms
of the symplectic form is presented in Section \ref{Sect6}. Some technical details and notation on symplectic
manifolds are collected in Section \ref{SymplMan}, while the summary
of our presentation is given in Section \ref{SecSumm}.
We set $\hbar = 1$ throughout the paper.

\section{Vectors, dual vectors and tensors} \label{SectAlgebra}

\noindent We shall start by recapitulating some basic
facts from the theory of vector spaces just to set up notation.
Standard references are \cite{Halmos,Axler,Hasani,AWH}. We consider only real
vector spaces here.

\subsection{Vectors and dual vectors}

Let $V$ be a vector space of dimension $N$. Each element $X \in V$
can be uniquely represented as $X = X^i e_i$, where $\{ e_i\}$ is
the set of basis vectors and $X^i$ are components of $X$ with respect
to this basis. Here, and throughout the paper, we employ
the Einstein summation convention so that in all expressions which contain the same letter as a lower and as an upper index,  the summation is carried over all the values of index: $1,2,3,...,N$.
Real $N$ dimensional vector space is isomorphic to $\mathbb{R}^N$, the space of  $N-$tuplets of real
numbers. By convention, we write these elements as columns. For example,
when $N = 2$, to each vector $X = X^1 e_1 + X^2 e_2$ we ascribe the column
\begin{equation}
    \begin{bmatrix}
       X^1     \\
      X^2         \\
		\end{bmatrix},
\end{equation}
which we denote by the same symbol, $X$.

Dual vectors are linear maps from $V$ to real numbers and,
as it turns out, they form a vector space of the same dimension
$N$. We denote this space by $V^*$ and a typical dual vector
$\alpha \in V^*$ is written as $\alpha = \alpha_i e^{*i}$. Here $\{ e^{*i}\}$ denotes the dual basis whose elements are defined by
their action on the basis elements of $V$, $e^{*i}(e_j) = \delta^i_j$.
Dual vectors are represented by rows. If $N=2$, we have
\begin{equation}
    \alpha = \alpha_1e^{*1} + \alpha_2 e^{*2} \equiv \begin{bmatrix} \alpha_1\;
    \alpha_2 \end{bmatrix}
\end{equation}
so that the action of $\alpha \in V^*$ on $X \in V$ can be written as
\begin{equation}
    \alpha(X) = \begin{bmatrix} \alpha_1\;
    \alpha_2 \end{bmatrix} \begin{bmatrix} X^1  \\ X^2  \\
		\end{bmatrix}
  = \alpha_1 X^1 + \alpha_2 X^2 \equiv X(\alpha),
\end{equation}
which is just a real number. 
The last step in the previous equation simply states that vectors
could be considered as linear maps on dual vectors.
Although we are usually a bit
cavalier when it comes to  vectors
and dual vectors, we must keep in mind that these are 
different mathematical objects that should be carefully distinguished. 

\subsection{Tensors}

Vectors and dual vectors can be used to construct more complicated
objects -- tensors. These are multilinear maps (i.e. linear with respect
to each argument) which take certain number of dual vectors and vectors
to produce a number. Formally, we say that the multilinear map
\begin{equation}
    {\rm T}: \underbrace{V^* \times V^* \times \dots \times V^*}_{{\rm n} \;\;\rm{copies}} \times \underbrace{V \times V \times \dots \times V}_{{\rm m} \;\; \rm{copies}} \to \mathbb{R},
\end{equation}
where $\times$ denotes Cartesian product, is a tensor of type $(\rm n,m)$. 
A tensor of type $(0,0)$ is a scalar, while those of type $(\rm n,0)$ are called
contravariant tensors of rank $\rm n$, while those of type $(0, \rm m)$ are
covariant tensors of rank $\rm m$.
We shall restrict ourselves here to 
tensors of types $(1,1)$, $(0,2)$ and $(2,0)$. 

Tensors of various types are naturally expressed in terms of tensor products.
Given two dual vectors, $\alpha, \beta \in V^*$, we define their
tensor product $\alpha \otimes \beta$ by
\begin{equation}
    \alpha \otimes \beta(X,Y) := \alpha(X) \beta(Y),
\end{equation}
where $X,Y \in V$. As dual vectors are linear maps, this construction automatically
satisfies the definition of a $(0,2)$ type tensor. In terms of its components ${\rm T}_{ij}$,
an arbitrary $(0,2)$ type tensor is written as ${\rm T} = {\rm T}_{ij} e^{*i} \otimes e^{*j}$, where ${\rm T}_{ij} = {\rm T}(e_i,e_j)$. In complete analogy, $(2,0)$ and $(1,1)$ tensors are given by
${\rm A}={\rm A}^{ij} e_{i} \otimes e_{j}$ and ${\rm B} = {\rm B}^i_{\;j} e_{i} \otimes e^{*j}$,
with components given by ${\rm A}^{ij} = {\rm A}(e^{*i}, e^{*j})$ and
${\rm B}^i_{\;j} = {\rm B}(e^{*i},e_j)$. This construction can be applied to tensors of arbitrary type.
In particular, a vector is a tensor of type $(1,0)$ and a dual vector is a type $(0,1)$ tensor. In physics literature, vectors and dual vectors are frequently labeled as contravariant and covariant vectors.
If ${\rm T}_{ij}
= {\rm T}_{ji}$, we say that the corresponding tensor is symmetric. On the other hand, if ${\rm T}_{ij}= -{\rm T}_{ji}$, the tensor is antisymmetric.

Tensor products of vectors and dual vectors could also be represented
by rows and columns. This operation is defined  so that each component of the
first factor in the product is multiplied by the entire second object (the object being a row or a column). For example,
\begin{equation}
e_1 \otimes e^{*1} = 
    \begin{bmatrix} 1  \\ 0  \\
		\end{bmatrix}
  \otimes \begin{bmatrix} 1\; 0 \end{bmatrix}
  = \begin{bmatrix} 1 \cdot  \begin{bmatrix} 1\; 0 \end{bmatrix}   \\[0.3em]
  0 \cdot \begin{bmatrix} 1\; 0 \end{bmatrix}  \\[0.3em]
		\end{bmatrix} = \begin{bmatrix}
       1 & 0    \\
       0 & 0        \\
     \end{bmatrix},
\end{equation}
or
\begin{align}
    e^{*1} \otimes e^{*2}& = \begin{bmatrix} 1\; 0 \end{bmatrix} \otimes
    \begin{bmatrix} 0\; 1 \end{bmatrix} = \begin{bmatrix} 1 \cdot \begin{bmatrix} 0\; 1 \end{bmatrix}
    \; 0 \cdot \begin{bmatrix} 0\; 1 \end{bmatrix} \end{bmatrix} \nonumber\\ 
    & = \begin{bmatrix} 0\; 1 \; 0 \; 0 \end{bmatrix}.
\end{align}
Thus, for a two-dimensional vector space,  a $(1,1)$ type tensor ${\rm B}$ is represented by a matrix
\begin{equation}
   {\rm B} =  {\rm B}^i_{\;j} e_{i} \otimes e^{*j} = \begin{bmatrix}
       {\rm B}^1_{\;1} & {\rm B}^1_{\;2}    \\[0.35em]
       {\rm B}^2_{\;1} & {\rm B}^2_{\;2}        \\
     \end{bmatrix},
\end{equation}
while  a $(0,2)$ type tensor is represented by
\begin{equation}
   {\rm T} =  {\rm T}_{ij} e^{*i} \otimes e^{*j} = 
   \begin{bmatrix} {\rm T}_{11}\; {\rm T}_{12} \; {\rm T}_{21} \; {\rm T}_{22}  \end{bmatrix}.
\end{equation}
An explicit representation of a $(2,0)$ tensor can be found in a similar
fashion. 

Components of tensors could as well be calculated using matrix notation.
For example,
\begin{align}
    {\rm B}^1_{\;2} = {\rm B}\left(e^{*1},e_2 \right) = 
    \begin{bmatrix} 1\;\;\; 0 \end{bmatrix}
    \begin{bmatrix}
       {\rm B}^1_{\;1} & {\rm B}^1_{\;2}    \\[0.35em]
       {\rm B}^2_{\;1} & {\rm B}^2_{\;2} 
       \end{bmatrix}
        \begin{bmatrix} 0  \\[4pt]    1  \\
		\end{bmatrix}
\end{align}
with analogous expressions in other cases.

\subsection{Metric tensor}  \label{MetricTensorSection}

A non-degenerate symmetric tensor of type $(0,2)$ is of particular significance. 
It goes by the name of metric tensor and, as it
is well known, it defines an inner product on $V$. If $\rm G$ is such a symmetric tensor and $X,Y \in V$,
their inner product is defined to be
\begin{equation}
    \langle X, Y \rangle := {\rm G}(X,Y) = {\rm G}_{ij}X^i Y^j.
\end{equation}
We can use elements ${\rm G}_{ij}$ to define the $(2,0)$ tensor
${\rm G}^{-1} \equiv {\rm G}^{ij}e_i \otimes e_j$, where ${\rm G}^{ij}$ are solutions
of equations ${\rm G}_{ik}{\rm G}^{kj} = \delta^j_{\;i}$.
The metric tensor also establishes an isomorphism between $V$ and $V^*$. 
Let $Y\in V$ be an arbitrary vector. Since ${\rm G}$ is symmetric, the dual vector
${\rm G}(Y, \rule{0.4cm}{0.15mm}) = {\rm G}(\rule{0.4cm}{0.15mm},Y)$
has the property that its action on a vector $X$ coincides with
the inner product of $X$ and $Y$. By convention, the components of this
dual vector are  denoted by $Y_i$. Thus,
\begin{equation}
       Y_i = {\rm G}_{ij}Y^j.
\end{equation}
The procedure of ascribing a dual vector to the vector $Y$
is known as lowering the index. Similarly, we can "raise the index" of
a dual vector $\alpha$ with the action of ${\rm G}^{-1}$ as ${\rm G}^{-1}(\alpha, \rule{0.4cm}{0.15mm}) = {\rm G}^{-1}(\rule{0.4cm}{0.15mm}, \alpha)$, which, in components, reads
\begin{equation}
    \alpha^i = {\rm G}^{ij} \alpha_j.
\end{equation}
This conversion between upper and lower indices works for tensors
of arbitrary type as well. For example
\begin{equation}
    {\rm T}^i_{\;jk} ={\rm G}^{il} {\rm G}_{jm}{\rm G}_{kn} {\rm T}_l^{\;mn}
\end{equation}
with summation over $l,m$ and $n$ understood.

By the standard abuse of notation, metric tensor is usually represented as
a square matrix with the elements ${\rm G}_{ij}$. With this convention,
${\rm G}^{-1}$ corresponds to the inverse matrix of $\rm G$, i.e.
$\rm G \rm G^{-1} = \rm I$, where $\rm I$ denotes $N \times N$ unit matrix.

\subsection{Linear operators}

A second class of tensors, which turns out to be of
special importance, are tensors of type $(1,1)$, also
known as linear operators. These tensors have a property to map
vectors onto vectors and dual vectors onto dual vectors.
Let ${\rm U} = {\rm U}^i_{\;j} e_i \otimes e^{*j}$
be such a tensor and it act on vectors in $\mathbb{R}^N$.
If $x \in \mathbb{R}^N$, then $\Bar{x} = {\rm U}(x)$ and
\begin{equation}
    {\Bar x}^i = {\rm U}^{i}_{\;j} x^j.
\end{equation}
This equation allows us to express matrix elements of ${\rm U}$ as \cite{RyderGR}
\begin{equation}
    {\rm U}^i_{\;j} = \frac{\partial {\Bar x}^i}{\partial x^j} \label{VectUDef}
\end{equation}
In this manner, we arrive at an alternate definition of a vector:
If a quantity $X$, whose components  $X^i$ transform with tensor
${\rm U}$ in such a way that
\begin{equation}
    {\Bar X}^i = \frac{\partial {\Bar x}^i}{\partial x^j} X^j \label{VecTransDef},
\end{equation}
we say that $X$ is a vector. The notation introduced in  \eqref{VectUDef} suggests that 
components of $X$ may depend on $x$. This will be the case when we introduce
vector fields.

Further, let ${\rm V} = {\rm V}^i_{\;j} e_i \otimes e^{*j}$ be a linear
operator which performs the corresponding transformation on dual vectors,
${\Bar \alpha} = {\rm V}(\alpha)$. Thus,
\begin{equation}
    {\Bar \alpha}_i = {\rm V}^j_{\; i} \alpha_j.
\end{equation}
The coefficients ${\rm V}^j_{\; i}$ cannot be arbitrary. If we assume
that the action of $\alpha$ on $X$ remains invariant, 
${\Bar \alpha}(\Bar X) = \alpha (X)$, then
\begin{equation}
    {\rm V}^k_{\;i}{\rm U}^i_{\;j} \alpha_k X^j = \alpha_k X^k
\end{equation}
gives conditions on components of ${\rm V}$, 
\begin{equation}
{\rm V}^k_{\;i}{\rm U}^i_{\;j} = \delta^k_{j} \label{UVinverse}
\end{equation}
and
we see from \eqref{VectUDef} and \eqref{UVinverse}  that
\begin{equation}
    {\rm V}^i_{\;j} = \frac{\partial {x}^i}{\partial {\Bar x}^j}.
    \label{DualVDef}
\end{equation}
A simple application of the chain rule proves the correctness of (\ref{DualVDef}). We can now say that a dual vector is a $N$ component
quantity $\alpha$ whose components transform according to
\begin{equation}
    {\Bar \alpha}_i = \frac{\partial {x}^j}{\partial {\Bar x}^i} \alpha_j.
    \label{DualTransfDef}
\end{equation}
One can easily derive transformation rules for tensors of higher order.
We shall have no use of them here, however.

On the other hand, we may use the tensor $\rm U$ to transform basis set
of $\{ e_i \}$ into $\{ E_i = {\rm U}(e_i)  \}$. The corresponding dual
basis is then given by $\{ E^{*i} = {\rm V}(e^{*i})  \}$ if 
${\rm V}^k_{\;i}{\rm U}^i_{\;j} = \delta^k_{j}$.

\subsection{Tensor fields}

At this point we are ready to generalize the algebraic concept of tensors
to objects which continuously depend on a certain number of parameters.
These are known as tensor fields and include vector and dual vector fields
as special cases. The concept of a tensor field not only allows us to differentiate or integrate proper objects, but is also  a  fundamental object
in non-Euclidean geometry. Detailed accounts of tensor fields and their application to physical problems can be found in
\cite{Fecko,Frankel}. We shall focus on the concept of gradient here.
In Euclidean geometry, the gradient of a scalar field $\phi$ is usually treated as a vector field with
components $\partial \phi/\partial x_i$ \cite{AWH}. However, if we change the coordinate
system and the express gradient in $\{ \Bar{x}^i\}$ coordinates, by the use of the chain rule, we find
\begin{equation}
  \frac{\partial \phi}{\partial { x}^i}  \longrightarrow  \frac{\partial \phi}{\partial {\Bar x}^i} = 
   \frac{\partial { x}^j}{\partial \Bar x^i}  \frac{\partial \phi}{\partial {x}^j}.
\end{equation}
According to \eqref{DualTransfDef}, the components of a gradient transform as components of a dual vector. Thus, in general, the gradient is a dual vector. This
object is usually denoted by $\d \phi$ since its components are the same as 
components of a total differential $\d \phi$ of a function $\phi$.
The fact that gradient is not truly a vector will be of crucial importance for obtaining the true form of Hamilton equations
in the case when phase space is a non-Euclidean space.

We also mention that metric tensor becomes a field which defines an inner product at each point of (in general, a non-Euclidean) space.

\section{Symplectic form on $\mathbb{R}^2$} \label{Sect3}

\noindent Suppose we have a two-dimensional phase space ${\rm Z} = \mathbb{R}^2$
with coordinates $(q,p)$ and a Hamiltonian function $H(q,p)$. The dynamics
of the system described by $H$ follows from the Hamilton equations
\begin{equation}
    \dot q = \frac{\partial H}{\partial p} ,\hspace{1cm}\dot p = - \frac{\partial H}{\partial q}. \label{HamEqCoord}
\end{equation}
We wish to generalize these equations to the phase space
which is ${\rm S}^2$. This can be done by casting them into a coordinate-free form. In this way we shall identify the  underlying  geometric structure carried by the so-called
symplectic form on $\mathbb{R}^2$. The geometric meaning of the symplectic form on
$\mathbb{R}^2$ will also give us a hint on how to construct
the corresponding tensor field on ${\rm S}^2$.

For a start, we introduce a vector field $X \in \mathbb{R}^2$
\begin{equation}
    X = \begin{bmatrix}
      q     \\
      p         \\
		\end{bmatrix}
\end{equation}
and the gradient of $H$ with respect to $X$
\begin{equation}
    \nabla_X H = \left[\begin{array}{cc}
       \frac{\partial H}{\partial q}  \\
       \frac{\partial H}{\partial p}  \\
      \end{array} \right] . \label{HamGradVec}
\end{equation}
To express \eqref{HamEqCoord} in coordinate-free form, we need to permute 
the components of the gradient vector \eqref{HamGradVec}. This could
be done with the help of the matrix
\begin{equation}
    \rm A = \left[\begin{array}{rr}
        0 & 1 \\
        -1 & 0
    \end{array} \right]. \label{OmegaR2Def}
\end{equation}
We can now write the system \eqref{HamEqCoord} as a single  equation
for $X$:
\begin{equation}
    \dot X = 
   {\rm A} \left(\nabla_X H \right). \label{HamEqFirst}
\end{equation}
The notation indicates that $\rm A$, as a matrix, acts on a column vector $\nabla_X H$
from the left. Equation \eqref{HamEqFirst} looks like a true vector equation, but to make sure this is indeed the case, we need to check whether the components of \eqref{HamEqFirst} satisfy the same relations in
all charts. Therefore, we change the basis vectors as
\begin{equation}
    e_i \to E_i = {\rm U}(e_i), \hspace{1cm}
    e^{*i} \to E^{*i} = {\rm V}(e^{*i})   \label{BasisChange}
\end{equation}
and
\begin{equation}
    \dot X = \dot X^i e_i \to {\rm U} (\dot X) = \dot X^k E_k = \dot X^k {\rm U}^l_{\;k} e_l \label{eq1}
\end{equation}
while, since $\mathbb{R}^2$ is an Euclidean space, we take $\nabla_X H$ to transform like a vector\footnote{Since there is no need to make a distinction between vectors and dual vectors in $\mathbb{R}^n$ with standard metric ${\rm G}_{ij} = \delta_{ij}$, all components of all quantities can be written with lower indices only. For example $X = X_i e_i$, $\omega = \omega_{ij} e_i \otimes e_j$ and so on.}
\begin{equation}
    \nabla_X H \to \frac{\partial H}{\partial X_k} {\rm U}^j_{\;k} e_j.
\end{equation}
Finally,
\begin{equation}
   {\rm A} \to {\rm A}^i_{j}{\rm U}^l_{\; i} {\rm V}^j_{\; m} e_{l} \otimes e^{*m},
\end{equation}
where ${\rm A}^i_{\;j}$ represent the components of the matrix \eqref{OmegaR2Def}.
Now
\begin{align}
    {\rm A}(\nabla_X H) &\to  {\rm A}^i_{j}{\rm U}^l_{\; i} {\rm V}^j_{\; m} e_{l} \otimes e^{*m}
    \left(  \frac{\partial H}{\partial X_k} {\rm U}^n_{\;k} e_n \right) \nonumber \\
    & = {\rm A}^i_{j}{\rm U}^l_{\; i} {\rm V}^j_{\; m}{\rm U}^m_{\;k}
    \frac{\partial H}{\partial X_k} e_{l} 
     = {\rm A}^i_{j}{\rm U}^l_{\; i} \frac{\partial H}{\partial X_j} e_{l} ,\label{eq2}
\end{align}
where we have used \eqref{UVinverse}.
By comparing \eqref{eq1} and \eqref{eq2}, we see that the components of \eqref{HamEqFirst} in the new frame satisfy 
\begin{equation}
    \dot{{X}}^i = {\rm A}^{i}_{\;j} \frac{\partial H}{\partial {X}^j}.
\end{equation}
Thus, the form of Hamilton equations \eqref{HamEqCoord} indeed remains the same.

However, equation \eqref{HamEqFirst} cannot be the whole story.
As we saw in the previous section, in general, gradient is not a vector.
This means that \eqref{HamEqFirst} is not an invariant equation in general.
Still, the matrix ${\rm A}$,  so-called  symplectic matrix, does have 
an intrinsic meaning on $\mathbb{R}^2$.
Namely,
it is used to measure the area enclosed by two vectors $X$ and $Y$. Indeed, by 
standard misuse of notation on Euclidean spaces, we have
\begin{align}
   X \cdot {\rm A} (Y)&= \begin{bmatrix}
        X^1 & X^2
    \end{bmatrix} \left[\begin{array}{rr}
        0 & 1 \\
        -1 & 0
    \end{array} \right] 
    \begin{bmatrix}
        Y^1 \\ Y^2
    \end{bmatrix} \nonumber \\
    &= X^1 Y^2 - X^2 Y^1 = {\rm Area}(X,Y).
\end{align}
We also mentioned in previous section that the quantity which produces
a number out of two vectors is a tensor of type $(0,2)$. In particular, 
the tensor in question is an antisymmetric tensor known as the symplectic form
\begin{equation}
    \omega = e^{*1}\otimes e^{*2} - e^{*2} \otimes e^{*1}
\end{equation}
so that
\begin{equation}
    \omega(X,Y) = {\rm Area} (X,Y).
\end{equation}
Thus, $\omega$ should replace $A$ in Hamilton equations. 

As a consistency check, we confirm that the action of $\omega$ on
$X,Y\in{\mathbb{R}}^2$ does not depend on the coordinates used. Since
\begin{align}
    \Bar \omega &= E^{*1}\otimes E^{*2} - E^{*2}\otimes E^{*1} \nonumber \\
    &= \left({\rm V}^1_{\;j} {\rm V}^2_{\;k} -
    {\rm V}^2_{\;j} {\rm V}^1_{\;k}\right) e^{*j}\otimes e^{*k},
\end{align}
we have
\begin{align}
    &\Bar \omega ({\Bar X},{\Bar Y})\nonumber  \\ = \left({\rm V}^1_{\;j} {\rm V}^2_{\;k} -
    {\rm V}^2_{\;j} {\rm V}^1_{\;k}\right) &X^l Y^i {\rm U}^m_{\;\;l}
    {\rm U}^n_{\;\;i} e^{*j}\otimes e^{*k} (e_m,e_n) \nonumber \\
    = X^1 Y^2 - X^2 &Y^1 = \omega (X,Y),
\end{align}
where we have used \eqref{BasisChange}. Note also that $\omega (X,X) = 0$,
while ${\rm G}(X,X) = |X|^2$.

If we want to write Hamilton equations with symplectic form,
a second issue arises. As we remarked earlier, in general, $\{\partial H/\partial X^i\}$ do
not form a vector, but a dual vector. 
To make a distinction between gradient vector $\nabla_X H$ and
this dual vector,
we shall denote the latter by $\d H$,
\begin{equation}
    \d H = \frac{\partial H}{\partial X^1} e^{*1} + \frac{\partial H}{\partial X^2} e^{*2}. \label{HamDif}
\end{equation}
Components of $\d H$ and $\nabla H$ are related by
\begin{equation}
(\nabla H)^i = {\rm G}^{ij}(\d H)_j,   \hspace{1cm}
    (\d H)_i = {\rm G}_{ij} (\nabla H)^j.
\end{equation}
Now we reach the final issue. Obviously, $\omega$ cannot
act on  $\nabla_X H$ to produce the vector $\dot X$. This must be done by a $(2,0)$ tensor. The tensor in question should
reduce to $A$ in Euclidean space so its components are
given by 
${\mathcal{P}}^{ik} \omega_{kj} = -\delta^i_j$.
Thus, if for $\dot X = \dot X^1 e_1 + \dot X^2 e_2$, the coordinate-free form of Hamilton equations is
\begin{equation}
    \dot X = {\mathcal{P}}(\rule{0.4cm}{0.15mm}, \d H),  \label{HamEqInv}
\end{equation}
where
\begin{equation}
    {\mathcal{P}} = {\mathcal P}^{ij} e_i \otimes e_j = \frac 12\mathcal{P}^{ij}\Big{(}e_i \otimes e_j - e_{j}\otimes e_i \Big{)}
\end{equation}
denotes the Poisson tensor -- an antisymmetric type $(2,0)$ tensor \cite{Fecko},
\begin{equation}
    \omega = \omega_{ij}e^{*i} \otimes e^{*j}  = \frac 12 \omega_{ij} 
    \Big{(}e^{*i} \otimes e^{*j} - e^{*j}\otimes e^{*i} \Big{)}
\end{equation}
and
\begin{equation}
    {\mathcal{P}}^{ik} \omega_{kj} = -\delta^i_j.   \label{PoissonSymplDef}
\end{equation}
Equation \eqref{HamEqInv}, besides being coordinate-free, is also suitable
for generalization to phase spaces other than $\mathbb{R}^2$.

We can use the Poisson tensor to write equations involving Poisson brackets
in coordinate-free form. Suppose $f$ is a function on phase space $\mathbb{R}^2$.
Then \cite{Witten}
\begin{equation}
    \dot f = \frac{\partial f}{\partial X^i} \dot{X}^i = 
   \frac{\partial f}{\partial X^i} {\mathcal{P}}^{ij} \frac{\partial H}{\partial X^j} \equiv \big{[}f, H \big{]}_{\rm{PB}}, \label{EOMPoiss}
\end{equation}
 and Poisson brackets for two functions on the phase space are defined as
\begin{equation}
    \big{[}f, g \big{]}_{\rm{PB}} := {\mathcal{P}} \Big{(} \d f, \d g \Big{)} ,\label{PBDef}
\end{equation}
 where $\d f = \partial_1 f e^{*1} + \partial_2 f e^{*2}$. 
 It is easy to see that the Poisson brackets defined in this way satisfy
 standard algebraic properties \cite{Fecko}. For example,
 \begin{align}
     \big{[}f, g h \big{]}_{\rm{PB}} & = {\mathcal{P}} \Big{(} \d f, \d (g h) \Big{)}\nonumber \\
     & = {\mathcal{P}} \Big{(} \d f, h \d g  \Big{)} + 
     {\mathcal{P}} \Big{(} \d f, g \d h   \Big{)} \nonumber \\
     & = h {\mathcal{P}} \Big{(} \d f,  \d g  \Big{)} + 
    g {\mathcal{P}} \Big{(} \d f, \d h   \Big{)} \nonumber \\
    & =h  \big{[}f, g  \big{]}_{\rm{PB}} + g \big{[}f, h  \big{]}_{\rm{PB}},
    \label{PoissonProduct}
 \end{align}
 where we have used Leibnitz rule for  $\d (f g)$ and linearity of
 the Poisson tensor.
 
 A pair of quantities on $\mathrm{R}^2$, which satisfies
\begin{equation}
    \big{[}f, g \big{]}_{\rm{PB}} = 1
\end{equation}
is known as the canonical pair of variables. In this simple
Cartesian space they are $f = X^1 = q$ and $g = X^2 = p$:
\begin{equation}
    \left[X^1, X^2 \right]_{\rm{PB}} = \big{[} q,p \big{]}_{\rm{PB}} = 1.
\end{equation}
The definition  \eqref{PBDef} is precisely the one which will give us the Poisson algebra of classical spins. A manifold ${\rm M}$ equipped with Poisson tensor $\mathcal{P}$ is known as a Poisson manifold \cite{Fecko}.

\section{Hamiltonian dynamics on two-sphere ${\rm S}^2$} \label{Sect4}

\subsection{Geometry of two-sphere}

\begin{figure}[t]
\begin{center}
\includegraphics[scale=0.5]{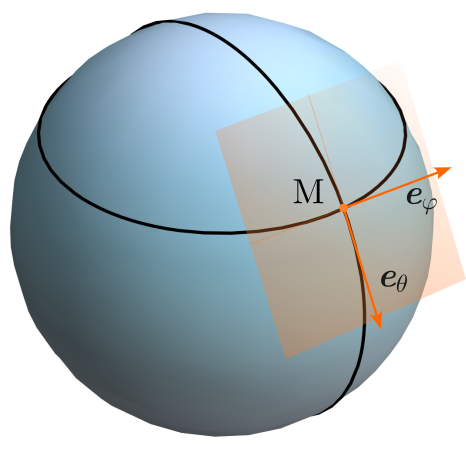}
\caption
{\label{fig1}Tangent plane to a sphere and two basis vectors $e_\varphi$ and 
$e_\theta$ at a point $\rm M \in S^2$.} 
\end{center}
\end{figure}

Our first task is to construct the symplectic form
on ${\rm S}^2$. To do so, we recall that, at each point ${\rm M} \in {\rm S}^2$, vectors
are defined as tangent vectors, i.e. as vectors lying in the tangent plane
at ${\rm M}$. To each point ${\rm M}$ we ascribe standard spherical coordinates
$\{\varphi, \theta\}$. The tangent plane is spanned by
vectors $\{e_{\varphi}, e_{\theta}\}$ and arbitrary vector field can be written
as $X(\varphi, \theta) = X^\varphi(\varphi,\theta)e_{\varphi} + 
X^\theta(\varphi,\theta)e_{\theta}$ [See FIG. \ref{fig1}]. The metric tensor on ${\rm S}^2$ is given by
\begin{equation}
    {\rm G} = \sin^2 \theta \; e^{*\varphi} \otimes e^{*\varphi} + e^{*\theta} \otimes e^{*\theta},
\end{equation}
where $\{e^{*\varphi}, e^{*\theta}\}$ represents the dual basis at each point 
of ${\rm S}^2$. That is, $e^{*\varphi}(e_{\varphi}) = e^{*\theta}(e_{\theta}) =1$.
Note that $\{e_{\varphi}, e_{\theta}\}$ is not an orthonormal basis since
$|e_{\varphi}|^2 = {\rm G}(e_{\varphi}, e_{\varphi}) = \sin^2 \theta$. This means that $e_\varphi$ corresponds to $\sin \theta \bm e_{\varphi}$,
where $\bm e_\varphi = - \sin \varphi \bm e_x + \cos \varphi \bm e_y$ is the standard unit vector field of the spherical coordinate
system.
Next, we recall the expression for an area element
on a unit sphere. In standard spherical coordinates $\{\varphi, \theta\}$, it is given by $\Delta A = \sin \theta \Delta \varphi \Delta \theta$ \cite{AWH}.
We interpret this result as an area enclosed by vectors
$\Delta \varphi e_{\varphi}$ and $\Delta \theta e_\theta$. 
Now we see that the symplectic form on ${\rm S}^2$ is given by \cite{MechSymm}
\begin{equation}
    \omega = - \sin \theta \Big{(}    e^{*\varphi} \otimes  e^{*\theta}-e^{*\theta} \otimes  e^{*\varphi} 
    \Big{)}    \label{SimpOmegaDef}
\end{equation}
so that
\begin{equation}
    \omega \Big{(}\Delta \theta e_\theta, \Delta \varphi e_{\varphi} \Big{)} = \Delta A.
\end{equation}
Therefore, the Poisson tensor field is given by
\begin{equation}
    {\mathcal P} = - \left( \sin \theta \right)^{-1}  \Big{(}  e_{\varphi} \otimes  e_{\theta}
    -  e_{\theta} \otimes  e_{\varphi}\Big{)}.  \label{PoissTenDef}
\end{equation}
Having the Poisson tensor at our disposal, we can calculate Poisson brackets and identify canonical variables. Let us 
try first with the most obvious choice $X^1 = \varphi, X^2 = \theta$.
Since $\d \varphi = e^{* \varphi}$ and $\d \theta = e^{* \theta}$, we have
\begin{equation}
    \big{[} \varphi,  \theta \big{]}_{\rm PB} = {\mathcal P} 
     \Big{(}   \d \varphi, \d \theta \Big{)} = 
     - \frac{1}{\sin \theta}.
\end{equation}
Thus, $\varphi$ and $\theta$
are not the canonical variables for ${\rm S}^2$. For our second guess, we take
$X^1 = \varphi, X^2 = \cos \theta$, with $\d \cos \theta = -\sin \theta e^{*\theta}$.
Since 
\begin{equation}
    \big{[} \varphi,  \cos \theta \big{]}_{\rm PB} = {\mathcal P} 
     \Big{(}   \d \varphi, \d \cos \theta \Big{)} = 1,   \label{PoissonBracketsSingle}
\end{equation}
we can select $\{ \varphi, \cos \theta  \} \equiv \{ q, p  \}$ to be a pair of canonical variables
on ${\rm S}^2$. This will allow us to define Hamiltonian systems on ${\rm Z} = {\rm S}^2$ and to study
their dynamics. In particular, we shall define classical spins and classical
Heisenberg model.

\subsection{A simple quadratic model on ${\rm S}^{2}$} \label{ToyS2}

As a warm-up, let us consider a simple quadratic model on the sphere ${\rm S}^2$,
with Hamiltonian defined as the direct generalization of linear harmonic oscillator
\begin{equation}
    H = \frac{p^2}{2m} + \frac{m \Omega^2 q^2}{2}
\end{equation}
with $q = \varphi$ and $p = \cos \theta$. To simplify notation, we choose
$m = \Omega = 1$ so that the quadratic Hamiltonian reduces to
\begin{equation}
    H = \frac{\cos^2 \theta}{2} + \frac{\varphi^2}{2}.   \label{S2QuadrHam}
\end{equation}
Corresponding dual vector field $\d H$ is
\begin{equation}
    \d H = -\sin \theta \cos \theta e^{* \theta} + \varphi e^{*\varphi}
\end{equation}
and the equations of motion \eqref{EOMPoiss} become
\begin{equation}
    \frac{\d \varphi}{\d t} = \cos \theta, \hspace{1cm} \frac{\d}{\d t} \cos \theta = 
    -\varphi. \label{LHOSystem}
\end{equation}
To solve this system, we apply the standard procedure and take
the second derivative w.r.t time
\begin{equation}
    \ddot \varphi = \frac{\d}{\d t} \dot \varphi = {\mathcal{P}}
    \Big{(} d \dot \varphi, \d H\Big{)}  = -\varphi,
\end{equation}
where $\d \dot \varphi = \d \cos \theta = -\sin \theta e^{*\theta}$.
By choosing $\varphi(0) = \varphi_0$ and $\dot \varphi (0)=0$,
we have
\begin{equation}
    \varphi(t) = \varphi_0 \cos t.  \label{LHOPhiSol}
\end{equation}
Upon  substituting $\varphi(t)$ into the second equation of \eqref{LHOSystem}
gives
\begin{equation}
    -\dot \theta \sin \theta = - \varphi_0 \cos t,
\end{equation}
so that
\begin{equation}
    \cos \Big{(} \theta(t) \Big{)} = p(t) = \cos \theta_0 - \varphi_0 \sin t,  \label{LHOThetaSol}
\end{equation}
where $\theta_0$ represents the initial value for $\theta(t)$.

\begin{figure}[t]
\begin{center}
\includegraphics[width=\columnwidth]{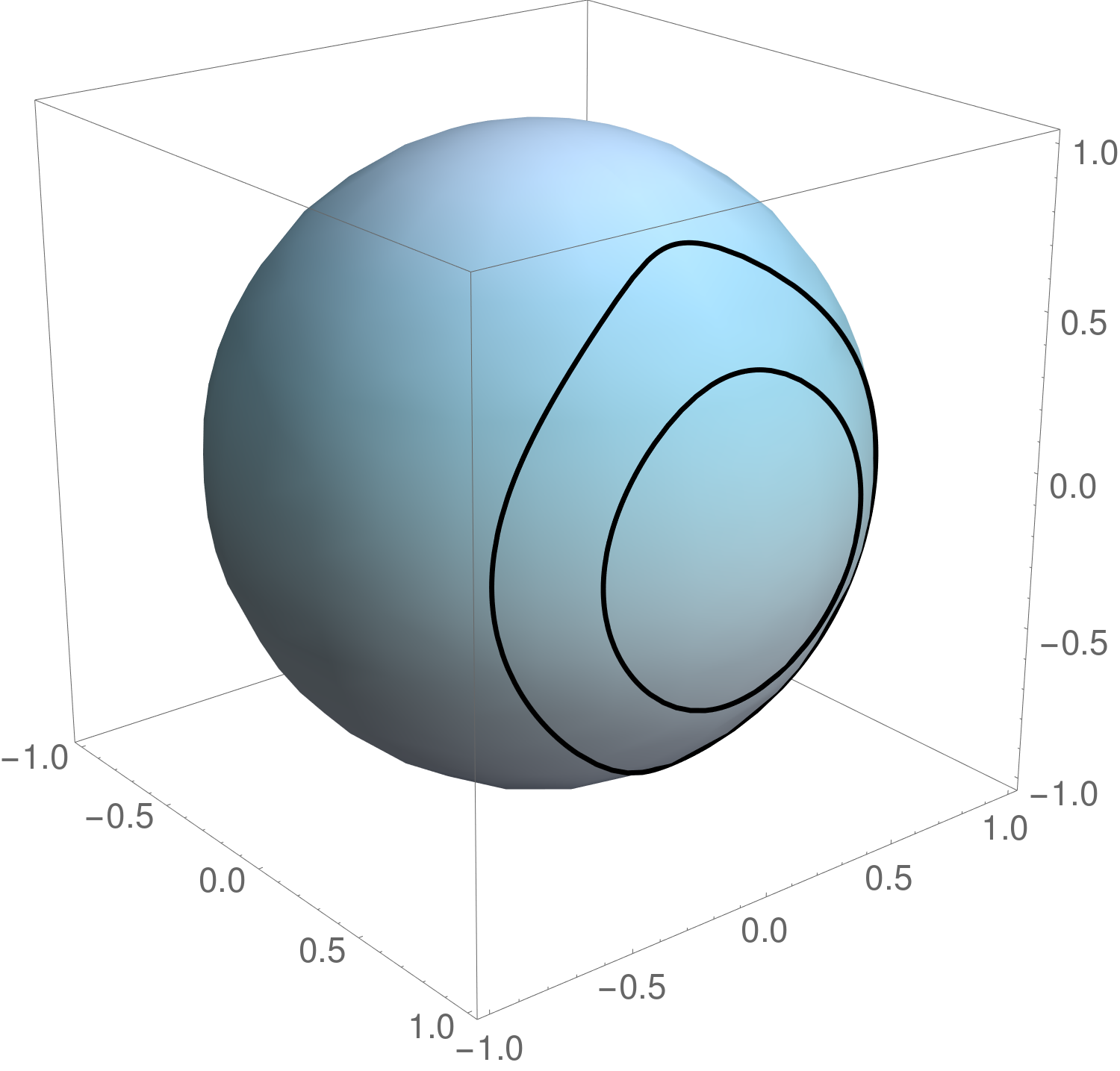}
\caption
{\label{fig2}Phase space trajectory $(\varphi(t), p(t) = \cos \theta(t))$ for a simple
quadratic Hamiltonian \eqref{S2QuadrHam} and two sets of initial conditions
$(\varphi_0 = 0.5, \cos \theta_0 = 0)$ and $(\varphi_0 = 0.8, \cos \theta_0 = 0)$.}
\end{center}
\end{figure}

\begin{figure}[t]
\begin{center}
\includegraphics[width=\columnwidth]{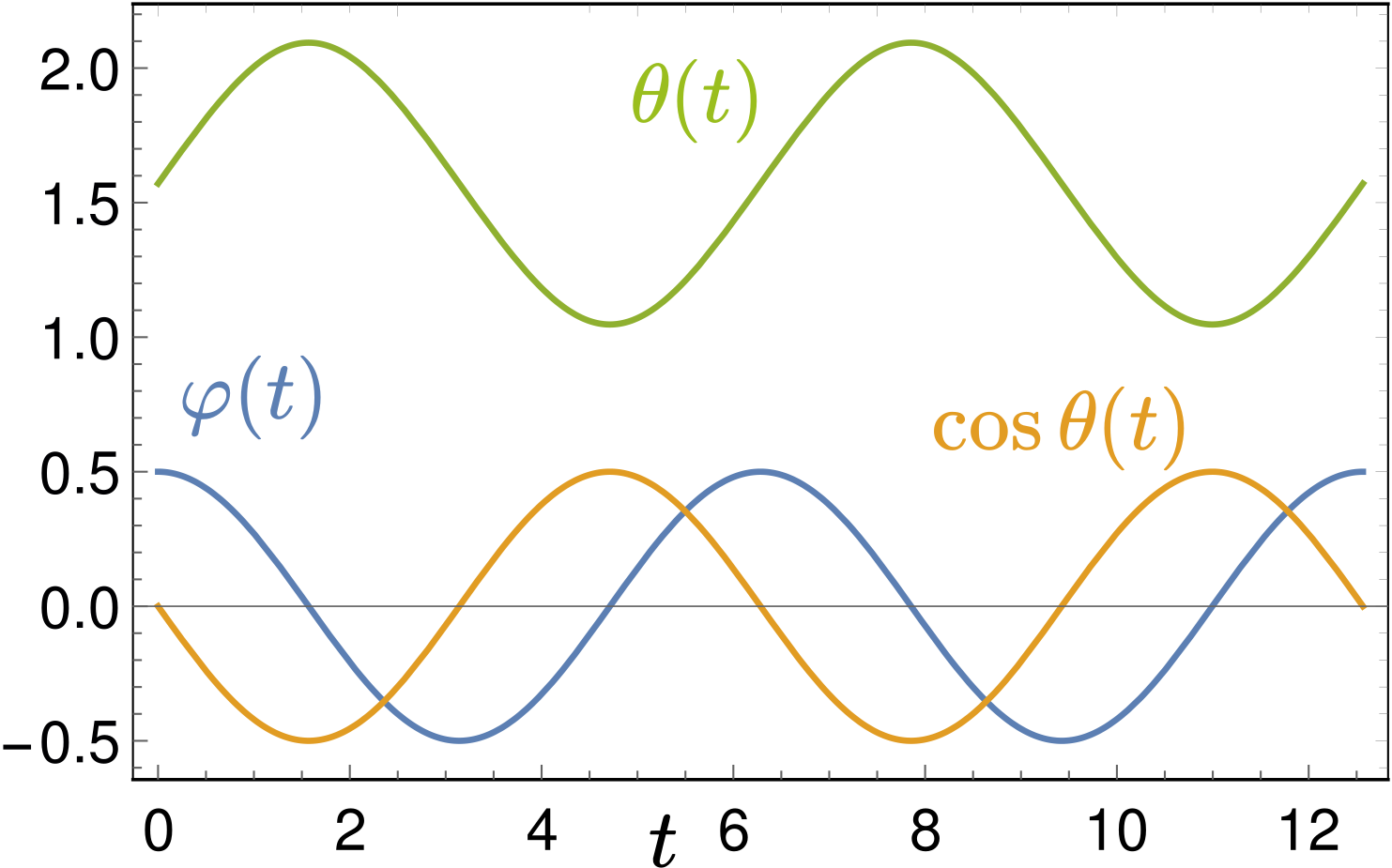}
\caption
{\label{fig3} Time evolution of phase space variables $\varphi(t), \theta(t)$ and $p(t)
= \cos \theta (t)$ for initial conditions
$(\varphi_0 = 0.5, \cos \theta_0 = 0)$.}
\end{center}
\end{figure}

Equations \eqref{LHOPhiSol} and \eqref{LHOThetaSol} constitute a  solution
to this simple quadratic model on ${\rm S}^2$ and $(\varphi(t), p(t))$ is a 
parametrized trajectory on two-sphere as a phase space. This solution
is presented in FIG. \ref{fig2}, while FIG. \ref{fig3} shows explicit
solutions $\theta(t), \varphi(t)$ and $p(t) = \cos \big{(}\theta (t) \big{)}$. As expected
from the Hamiltonian \eqref{S2QuadrHam}, phase-space trajectories resemble
those of a simple linear harmonic oscillator.
Note that the model \eqref{S2QuadrHam} does not describe
a particle confined to a sphere. For a geometric treatment
of classical and quantum particle on a sphere, see \cite{GuerreroS2, SilvaJPA}.

\section{Classical Heisenberg model} \label{ClassHeisSect}

\noindent
Now that we have defined the symplectic form and the Poisson tensor,
and gained some insight into the phase space ${\rm S}^2$ by analyzing a simple
harmonic model, we proceed 
with the study of classical spins.

\subsection{Poisson algebra of classical spins}

We shall first focus on the case of a single classical spin \cite{Stone}. 
A classical unit vector can be represented as
\begin{equation}
    \bm S =  \begin{bmatrix}
       S_x     \\
      S_y         \\
     S_z \\
		\end{bmatrix}
    =
    \begin{bmatrix}
       \sin \theta \cos \varphi     \\
      \sin \theta \sin \varphi         \\
      \cos \theta \\
		\end{bmatrix} \label{ClassSPinComponents}
\end{equation}
and the condition $|\bm S| = 1$ follows directly from \eqref{ClassSPinComponents}.
Since any state of $\bm S$ can be uniquely mapped onto a point
on a unit sphere, we can consider $\bm S$ to be a dynamical variable
defined in terms of the canonical pair $\{\varphi, \cos \theta\}$ which
parametrizes the phase space ${\rm S}^2$. To specify the dynamics, we need
a Hamiltonian. However, as we have identified the underlying phase space
and corresponding Poisson tensor, we can derive expressions for Poisson
brackets of $S_x, S_y$ and $S_z$. Note that $S_x, S_y$ and $S_z$ are
not components of a vector (or a vector field) on ${\rm S}^2$. As far as the geometry of phase
space is concerned, they are dynamical variables built out of canonical
coordinates $\{\varphi, \cos \theta\}$. Because of that, we choose to
write  $x,y$ and $z$ as lower indices.

Let us calculate first the Poisson bracket of $S_x$ and $S_y$. According to
\eqref{PBDef}, we have
\begin{equation}
    \Big{[} S_x, S_y \Big{]}_{\rm PB} = \mathcal{P} \Big{(} \d S_x, \d S_y  \Big{)},
\end{equation}
with ${\mathcal P}$ given in \eqref{PoissTenDef} and
\begin{equation}
\begin{split}
    \d S_x & = - \sin\theta \sin \varphi e^{*\varphi} + \cos \theta \cos \varphi e^{*\theta}, \\
    \d S_y & = \sin\theta \cos \varphi e^{*\varphi} + \cos \theta \sin \varphi e^{*\theta},\\
    \d S_z & = -\sin \theta e^{*\vartheta}
\end{split}
\end{equation}
Therefore,
\begin{equation}
\begin{split}
     \Big{[} S_x, S_y \Big{]}_{\rm PB} &= \sin\theta \sin \varphi \frac{\cos \theta}{\sin \theta} \sin \varphi + \cos \theta \cos^2 \varphi \\
     & = \cos \theta = S^z. \label{PZ1}
\end{split}
\end{equation}
In a similar manner, we find
\begin{align}
    \Big{[} S_y, S_z \Big{]}_{\rm PB} & = \mathcal{P} \Big{(} \d S_y, \d S_z  \Big{)}
    = \frac{1}{\sin \theta}  \sin \theta \cos \varphi \cdot (\sin \theta) \nonumber \\
   & = S_x \label{PZ2}
\end{align}
as well as
\begin{equation}
    \Big{[} S_z, S_x \Big{]}_{\rm PB} = S_y \label{PZ3}.
\end{equation}
By knowing that $\mathcal P$ is 
an antisymmetric tensor, we see that \eqref{PZ1}, \eqref{PZ2} and \eqref{PZ3} 
could be written using a single equation
\begin{equation}
     \Big{[} S_\alpha, S_\beta \Big{]}_{\rm PB} = \epsilon_{\alpha \beta \gamma} S^\gamma  \label{SingleSpinAlgebra}
\end{equation}
which represents the general form of Poisson brackets for classical
spins. Here $\epsilon_{\alpha \beta \gamma}$ represents Levi-Civita
symbol \cite{AWH}. 
Hence, the algebraic
relations under Poisson brackets which 
components of classical spins satisfy,  closely resemble the commutation relations
of spin operators \eqref{SpinCCR}.

\subsection{Hamiltonian and phase space}

The classical Heisenberg model is the simplest $\rm O(3)$
invariant model built using the phase space ${\rm S}^2$. 
To construct its Hamiltonian, we need to specify how points on ${\rm S}^2$
transform under the action of $\rm O(3)$.
The canonical
coordinates $\{ \varphi, \cos \theta  \}$ transform under  $\rm O(3)$ 
according to a nonlinear realization \cite{Burgess} and, because of that,
invariant Hamiltonian must include the metric tensor on ${\rm S}^2$. On the other
hand, the spin vector $\bm S$ transforms linearly and the Hamiltonian
takes a much simpler form when expressed in terms of $S_x, S_y$ and $S_z$.

For simplicity, let us assume a cubic lattice with sites labeled by
vector $\bm n$ and suppose further that each lattice site hosts a classical
spin $\bm S(\bm n)$. Since $|\bm S(\bm n)| =1$, we can build a Hamiltonian
function quadratic in $\bm S(\bm n)$ using projections of $\bm S(\bm n)$ on $\bm S(\bm m)$. If we assume that only nearest-neighbor projections contribute with
equal weights $J$, we arrive at the Hamiltonian of classical Heisenberg model
\begin{equation}
    H = - \frac{J}{2} \sum_{\bm n, \bm \lambda} \bm S(\bm n) \cdot \bm S(\bm n + \bm \lambda), \label{ClassHeisHam}
\end{equation}
where $\{ \bm \lambda\}$ denotes vectors connecting each site with its nearest 
neighbors. If $J>0$, the Hamiltonian \eqref{ClassHeisHam} will give rise to  ferromagnetic order \cite{Holm}.

Classical spin configurations satisfy equations of motion which can
be obtained using Poisson brackets for classical spins. Since Hamiltonian
\eqref{ClassHeisHam} describes a many-spin system, we need to generalize  the Poisson algebra \eqref{SingleSpinAlgebra}. The phase space of classical Heisenberg
model is given by Cartesian products of two-spheres 
for each lattice site
\begin{equation}
   {\rm{Z}} = \underset{\bm n}{\raisebox{-4.5pt}{\scalebox{2.8}{$\times$}}}  
   \left( {\rm S}^2 \right)_{\bm n} 
\end{equation}
and it represents  the space of solutions of classical equations of motion \cite{CrnkovicWitten}.
We shall denote spherical coordinates on a unit sphere located at lattice site $\bm n$ by
$\{ \varphi_{\bm n},  \theta_{\bm n}  \}$, 
the tangent vector fields to the unit sphere at $\bm n$
by $\{ e_{\varphi_{\bm n}}, e_{\theta_{\bm n}}  \}$
and the corresponding dual vector fields by $\{ e^{*\varphi_{\bm n}}, e^{*\theta_{\bm n}} \}$. The action of dual vector fields on $\{   e_{\varphi_{\bm n}}, e_{\theta_{\bm n}} \}$ is given by
\begin{equation}
     e^{*\varphi_{\bm n}} \Big{(} e_{\varphi_{\bm m}} \Big{)} = 
     e^{*\theta{\bm n}} \Big{(} e_{\theta{\bm m}} \Big{)} = \Delta(\bm n - \bm m),
\end{equation}
where $\Delta(\bm n - \bm m) =1$ only if $\bm n = \bm m$, $\Delta(\bm n - \bm m) = 0$ otherwise,
and the symplectic form generalizes to
\begin{equation}
    \omega = - \sum_{\bm n} \sin \theta_{\bm n}
     \Big{(}  e^{*\varphi_{\bm n}} \otimes  e^{*\theta_{\bm n}}
    -  e^{*\theta_{\bm n}} \otimes  e^{*\varphi_{\bm n}}\Big{)}. \label{SymplHeisMod}
\end{equation}
For each $\bm n$, a small variation of a  solution of equations of motion
can be written as $\delta \bm S(\bm n) = X^{\varphi} \bm e_{\varphi_{\bm n}}+X^{\bm \theta} \bm e_{\theta_{\bm n}}$ \cite{CrnkovicWitten}
so that $\bm S \cdot \delta \bm S = 0$ and $|\bm S + \delta \bm S| = 1$ within linear approximation in $\delta \bm S$.
Tensor field $\omega$ measures the area determined by two vector fields
in the sense that
\begin{equation}
    \omega \Big{(}\Delta \theta_{\bm m} e_{\theta_{\bm m}}, \Delta \varphi_{\bm n} e_{\varphi_{\bm n}} \Big{)} = \Delta(\bm n - \bm m)
    \sin \theta_{\bm m} \Delta \theta_{\bm m} \Delta \varphi_{\bm n}.
    \nonumber
\end{equation}
In complete analogy, we define the Poisson tensor
\begin{equation}
    {\mathcal{P}} = - \sum_{\bm n} \frac{1}{\sin \theta_{\bm n}}
    \Big{(}  e_{\varphi_{\bm n}} \otimes  e_{\theta_{\bm n}}
    -  e_{\theta_{\bm n}} \otimes  e_{\varphi_{\bm n}}\Big{)}
    \label{PoissTenSystDef}
\end{equation}
so that the Poisson brackets of spin components evaluate to
\begin{equation}
    \Big{[} S_\alpha(\bm n), S_\beta(\bm m) \Big{]}_{\rm PB} = \Delta(\bm n - \bm m)\varepsilon_{\alpha \beta \gamma} S^\gamma (\bm n) ,\label{LocalSpinAlgebra}
\end{equation}
while fundamental Poisson brackets \eqref{PoissonBracketsSingle} become
\begin{equation}
    \Big{[} \varphi_{\bm n}, \cos \theta_{\bm m} \Big{]}_{\rm PB} = \Delta(\bm n - \bm m).  \label{LocalPhaseAlgebra}
\end{equation}
Equation \eqref{LocalSpinAlgebra}, together with classical Hamiltonian
\eqref{ClassHeisHam}, completely specifies the dynamics of classical
spins in Heisenberg model. The Poisson bracket algebra \eqref{LocalPhaseAlgebra}
is usually derived starting from commutation relations for spin operators
expressed in terms of phase operator $\hat{\varphi}_{\bm n}$ and $\hat{S}^z_{\bm n}$
\cite{Villain,ClasLimit}, or postulated \cite{BraunerBook}, while relations \eqref{LocalPhaseAlgebra} are obtained by assuming
the correct form of the equations of motion for spins \cite{YangPB}. Here we see that both sets of algebraic relations arise due to symplectic structure
on $\rm Z$.
The transition to the theory of quantum   Heisenberg
 model 
can be achieved
by the standard canonical prescription: replace the classical variable $\bm S$ with
the operator $\hat{\bm S}$ at each lattice site $\bm n$ and replace the algebraic structure of Poisson brackets \eqref{LocalSpinAlgebra}
with commutators
\begin{equation}
    \Big{[} \;, \;\Big{]}_{\rm PB} \longrightarrow \frac{1}{\i}  \Big{[}\; ,\; \Big{]}.
\end{equation}
This line of reasoning allows us to arrive at the theory of quantum spin systems in a  manner which is
consistent with standard introduction of operator algebras in quantum systems --
by starting from the Poisson algebra of classical dynamical variables and
corresponding classical Hamilton function. Thus, the classical spin variable \eqref{ClassSPinComponents} defined
as a dynamical degree of freedom on phase space $\rm Z = S^2$ becomes an operator
\begin{equation}
    \bm S \longrightarrow \hat{\bm S}
\end{equation}
and the Poisson bracket algebra \eqref{SingleSpinAlgebra} becomes a set
of operator identities
\begin{equation}
\Big{[} S_\alpha, S_\beta \Big{]}_{\rm PB} = \epsilon_{\alpha \beta \gamma} S^\gamma
\longrightarrow \Big{[} \hat S_\alpha, \hat S_\beta \Big{]} = \i \epsilon_{\alpha \beta \gamma} \hat S^\gamma
\end{equation}
The extension to many-spin systems is now straightforward and readers can find details
in texts on Heisenberg and related models \cite{Auerbach,Nolting,Mattis,CMFT,Yosida}.

\subsection{Classical spin waves}

To make an analogy with classical field theories of scalar fields, it is useful
to rewrite Hamiltonian \eqref{ClassHeisHam} using the lattice Laplacian. 
In terms of a site-dependent quantity $\phi$, it is defined as
\cite{AnnPhys2015}
\begin{equation}
   \blacktriangledown^2 \phi(\bm x)   = \frac{2D}{Z_1 |\bm \lambda|^2} 
 \sum_{  \{ \bm \lambda \}  } \Big{[}\phi(\bm x + \bm \lambda) 
-  \phi(\bm x)\Big{]},
\label{OpIzLapA}
\end{equation}
where $Z_1$ denotes the number of nearest neighbors and $D$ is the 
dimension of the lattice. In case at hand, $D=3$ and $Z_1 = 6$.
A nice property of the lattice Laplacian is that its eigenfunctions
are plane waves
\begin{align}
    \blacktriangledown^2 \exp [\i \bm k \cdot \bm x] & =  
-\frac{2 D}{|\bm \lambda|^2} [1-\gamma_{D}(\bm k)] \exp [\i \bm k \cdot \bm x] 
\label{kSq} \nonumber\\
 &\equiv   - \mathcal{K}^2 \exp [\i \bm k \cdot \bm x] ,
 \end{align}
where
 \begin{equation}
 \gamma_{D}(\bm k)  =  Z_1^{-1} \sum_{\{ \bm \lambda \}} \exp [\i \bm k \cdot \bm \lambda].
\end{equation}
The lattice Laplacian is designed in such a way that
\begin{equation}
\lim_{|\bm \lambda| \to 0} \mathcal{K}^2 = \bm k^2 + \mathcal{O} (k^4)
\label{LatLaplCont}
\end{equation}
and, consequently,
\begin{equation}
    \blacktriangledown^2  \phi(\bm x) = \nabla^2 \phi(\bm x) + \mathcal{O}(|\bm \lambda|^2)
\label{LattLaplLimit}
\end{equation}
with $\nabla^2$ denoting standard Laplacian operator.  Hamiltonian \eqref{ClassHeisHam} can be rewritten using lattice Laplacian as
\begin{equation}
    H = 
 -\frac{1}{2}  \;\frac{J Z_1 |\bm \lambda|^2 }{2D} 
 \sum_{\bm n}  \bm S(\bm n)   \cdot 
\blacktriangledown ^2 \bm S(\bm n)
 -  \frac{J Z_1 N}{2}, \label{HFM2}
\end{equation}
where $N$ denotes the number of sites in the lattice.

Equations of motion
are now given by \eqref{EOMPoiss}, with  $f$ denoting components of $\bm S_{\bm n}$, $H$ given in \eqref{HFM2} and $\mathcal{P}$ defined in \eqref{PoissTenSystDef}. Thus,
\begin{align}
    \dot{\bm S}(\bm n,t) &= \Big{[} \bm S(\bm n,t) ,H   \Big{]}_{\rm PB}
    = \Big{[}  S_\beta(\bm n,t) ,H   \Big{]}_{\rm PB} \bm e_\beta \nonumber \\
    & = \frac{J Z_1 |\bm \lambda|^2 }{2D} \varepsilon_{\gamma \alpha \beta} S_{\gamma}(\bm n,t) \blacktriangledown^2 S_{\alpha}(\bm n,t) \bm e_{\beta}, \nonumber \\
    & = \frac{J Z_1 |\bm \lambda|^2 }{2D} \bm S(\bm n,t) \times \blacktriangledown^2 \bm S(\bm n,t), \label{LLE}
\end{align}
where we have used \eqref{LocalSpinAlgebra} and \eqref{PoissonProduct}
and the symbol $\times$ denotes the cross product. 
The equation \eqref{LLE} is a  discretized version of the Landau-Lifshitz equation \cite{LLreview}.

At this point we are going to make an approximation by assuming that spin
configurations governed by
\eqref{LLE} describe small deviations from a uniform $[0,0,1]^{\rm T} =  \bm e_z$ state.
Therefore, we take $\bm S$ to be of the form $[\delta S_x, \delta S_y, 1]^{\rm T}$, with $(\delta S_x)^2, (\delta S_y)^2 \ll 1$ and $\dot S_z = 0$. The equation \eqref{LLE} now reduces to
\begin{align}
    \dot{\delta S_x}(\bm n,t) & = -\frac{J Z_1 |\bm \lambda|^2 }{2D}  \blacktriangledown^2 S_y(\bm n,t), \nonumber \\
    \dot{\delta S_y}(\bm n,t) & = \frac{J Z_1 |\bm \lambda|^2 }{2D}  \blacktriangledown^2 S_x(\bm n,t).
\end{align}
As we shall see now, these two equations describe a single physical degree of freedom. This is most
easily seen by defining a complex field 
\begin{equation}
      \psi(\bm n,t) := \frac{\delta S_x(\bm n,t) + \i \delta S_y(\bm n,t)}{\sqrt{2}}
\end{equation}
so that the equation of motion for $\psi$ takes the  form of a Schr\"{o}dinger equation
\begin{equation}
    \i \dot{\psi}(\bm n,t) = - \frac{1}{2m}  \blacktriangledown^2
    \psi(\bm n,t), \label{MagnonEqn}
\end{equation}
where
\begin{equation}
    m:= \frac{D}{J Z_1 |\bm \lambda|^2}
\end{equation}
is a parameter that depends on the strength of the coupling 
$J$ and the structure of the lattice. The field $\psi$, which  describes small fluctuations around the
ferromagnetic ground state, is the classical spin wave. Being linear,
equation \eqref{MagnonEqn} can be solved by Fourier transform
\begin{equation}
    \hspace*{-0.4cm}\psi(\bm n,t) = 
\int_{\bm k} a_{\bm k}\; \e^{\i \bm k \cdot \bm x - \omega(\bm k) t},
\hspace{0.5cm}\int_{\bm k} \equiv  \int_{\rm{IBZ}} \frac{\d^D \bm k}{(2 \pi)^D}
\label{RPAPW}
\end{equation}
where
\begin{equation}
    \omega(\bm k) = \frac{{\mathcal K}^2}{2 m} \label{ClassFMDisp}
\end{equation}
and $a_{\bm k}$ are complex amplitudes and $\int_{\rm{IBZ}}$ denotes
integration over the first Brillouin zone, which is the volume of reciprocal space
containing all representative wave vectors \cite{CMFT}.

It is easy to see that the equation of motion \eqref{MagnonEqn}
follows from the Hamiltonian
\begin{equation}
    H_{\rm{LSW}} = -\frac{1}{2 m} \sum_{\bm n} \psi^*(\bm n,t) \blacktriangledown^2 \psi(\bm n,t)
\end{equation}
and Poisson brackets
\begin{equation}
    \Big{[} \psi(\bm n,t), \psi^*(\bm m,t)  \Big{]}_{\rm{PB}} = -\i \Delta(\bm n-\bm m),
\end{equation}
where 
\begin{equation}
    \Big{[} F, G \Big{]}_{\rm{PB}} = \sum_{\bm n} \left( \frac{\partial F}{\partial \psi(\bm n)} \frac{\partial G}{\partial \psi^*(\bm n)} - 
    \frac{\partial F}{\partial \psi^*(\bm n)} \frac{\partial G}{\partial \psi(\bm n)}
    \right). \nonumber
\end{equation}
This motivates us to define conjugate momentum $\pi:=\i \psi^*$, so that
\begin{equation}
    \Big{[} \psi(\bm n,t), \pi(\bm m,t)  \Big{]}_{\rm{PB}} = \Delta(\bm n-\bm m).
    \label{CanonnPBSch}
\end{equation}
Indeed, these all are well-known results in the theory of classical
Schr\"{o}dinger field \cite{Schiff}. The  symplectic
form of the linearized system is given by
\begin{equation}
    \omega = \sum_{\bm n} \Big{(} e^{* \psi_{\bm n}} \otimes e^{*\pi_{\bm n}}-  e^{*\pi_{\bm n}} \otimes e^{* \psi_{\bm n}} \Big{)} \label{SymplHeisModSW}
\end{equation}
and it describes a symplectic structure on the phase space $\mathbb{R}^{2 N}$, which
is the space of plane-wave solutions. In the context of classical spin waves, the  linearization process may be viewed as the
transition from \eqref{SymplHeisMod} to \eqref{SymplHeisModSW}.

To summarize, classical field configurations which describe small
deviations from the ferromagnetic ground state $\bm S (\bm n,t) = \bm e_z$
are given by (superposition of)  plane waves with dispersion relation
\eqref{ClassFMDisp}. In modern nomenclature, they are classified as
Nambu-Goldstone fields of type B \cite{PRX,ArXiv2024}.
Upon quantization, these will become elementary
excitations -- magnons.

We see from \eqref{LatLaplCont} and \eqref{ClassFMDisp} that in the
low-energy limit $\omega(\bm k) \approx \bm k^2/(2m)$,
so that ferromagnetic magnons behave like nonrelativistic particles.
This is not a universal feature of Nambu-Goldstone bosons
as evidenced by the cases of antiferromagnetic magnons, phonos or pions
which all exhibit relativistic dispersion relation $\omega (\bm k) \propto |\bm k|$ \cite{BraunerBook}. It turns out that nonrelativistic
dispersion relation is a consequence of the symplectic form defined
on the configuration space of magnon fields, which is also ${\rm S^2}$.
This symplectic form, also known as Kirillov-Kostant  symplectic form,
 is induced by spontaneous magnetization and it is 
     responsible for pairing two components $\{S_x(\bm x,t)$, $S_y(\bm x,t)\}$ into a single physical
 degree of freedom $\psi(\bm x,t)$ \cite{ArXiv2024}.

\section{Alternative definition of Poisson brackets} \label{Sect6}

\noindent As we have seen, the Poisson tensor can be
used to define Poisson brackets on a non-Euclidean phase space.
However, there is an equivalent definition in terms of the
symplectic form. Since the symplectic form is a  $(0,2)$ type
tensor field, it naturally acts on pair of vector fields and
these vector fields must be associated to the pair of functions
$f,g$ whose brackets we wish to define. 

We shall denote the vector field corresponding to the function $f$
by $X_f$ and define it in terms of $\d f$ \cite{Fecko}
\begin{equation}
    X_f = {\mathcal{P}}  \Big{(} \rule{0.4cm}{0.15mm} , \d f  \Big{)}.
    \label{XfDef}
\end{equation}
This definition resembles the usual index gymnastics described
in Section \ref{MetricTensorSection} except the fact that ${\mathcal P}$
is an antisymmetric rather than a symmetric tensor field.
The Poisson bracket is now given by \cite{MechSymm}
\begin{equation}
    \Big{[} f,g  \Big{]}_{\rm{PB}} := \omega  \Big{(} X_f, X_g  \Big{)},
    \label{PBOmega}
\end{equation}
where $f$ and $g$ are  functions on $\rm Z$.

Of course, we need to show that \eqref{PBOmega} coincides with 
\eqref{PBDef}. First of all, we see from \eqref{XfDef} that the components
of the associated vector field $X_f$ are given by $(X_f)^i = {\mathcal P}^{ik}\partial_k f$. Now
\begin{align}
    \omega  \Big{(} X_f, X_g  \Big{)} & = \omega_{ij} (X_f)^i (X_g)^j
    = \omega_{ij} {\mathcal P}^{ik} {\mathcal P}^{jl}  \partial_k f
    \partial_l g \nonumber \\
    & = -\omega_{ji}{\mathcal P}^{ik} {\mathcal P}^{jl}  \partial_k f
    \partial_l g \nonumber \\
    & = {\mathcal P} \Big{(} \d f, \d g  \Big{)} = 
    \Big{[} f,g  \Big{]}_{\rm{PB}},
\end{align}
where we have used \eqref{PoissonSymplDef}.
The two definitions are, therefore, completely equivalent and all results
regarding the phase space $\rm S^2$ and classical Heisenberg model could
be derived starting from \eqref{PBOmega}.

\section{Symplectic manifolds } \label{SymplMan}

\noindent The purpose of this section is to provide some details
on manifolds, and symplectic manifolds in particular.
The exposition is elementary and a complete treatment can be found
in \cite{Fecko,Arnold,Wood}.

First of all, a manifold is defined to be a space which locally  looks like
$\mathbb{R}^n$. For example, a sphere looks locally like $\mathbb{R}^2$
since the vicinity of each point on the sphere can be mapped on a two dimensional
Euclidean space by a one-to-one continuous map \cite{Fecko,Frankel}. The equivalence between $\rm S^2$ and $\mathbb{R}^2$ does
not hold globally, however.
Symplectic manifolds carry additional structure -- a closed non-degenerate
antisymmetric tensor field of type $(0,2)$. The condition of the symplectic form 
being closed amounts to  Poisson brackets  satisfying the
Jacobi identity. Non-degeneracy of an antisymmetric tensor field means that
we require non-degeneracy of symplectic form at each point on the manifold.
Given a point $\rm M$ of the manifold and a tangent vector to the manifold
at the same point $\rm M$, $X_{\rm M}$, the symplectic form at this
point, $\omega_{\rm M}$, is non-degenerate if $\omega_{\rm M}(X_{\rm M}, Y_{\rm M}) = 0$ implies $Y_{\rm M} = 0$. The condition of non-degeneracy seems to be
in conflict with our definition of symplectic form on $\rm S^2$ given in \eqref{SimpOmegaDef}.

To see that the symplectic form \eqref{SimpOmegaDef} is truly 
non-degenerate, we shall first show that it can be defined
in a coordinate-free manner. For a point $\rm{M} \in {\rm S}^2$, determined
by the unit vector $\bm e_r$, and two tangent vectors $X_{\rm M},Y_{\rm M}$ at
the same point, the symplectic form is defined by
\begin{equation}
   \omega_{\rm M}(X_{\rm M},Y_{\rm M}):= \bm e_r \cdot ( \bm X \times \bm Y  ) ,\label{OmegaProdDef}
\end{equation}
where $\bm X, \bm Y \in \mathbb{R}^3$ are  vectors tangent to the sphere at $\rm M$ (corresponding to $X_{\rm M}$ and $Y_{\rm M}$, respectively), $\times$ denotes  cross product of vectors in $\mathbb{R}^3$, and $\cdot$ is the standard inner
product on $\mathbb{R}^3$. The form $\omega$ is obviously well defined
at all points $\rm{M} \in {\rm S}^2$. 
The non-degeneracy condition for $\omega_{\rm M}$ is now easily established
since, given two nonzero and non-collinear vectors $X_{\rm M}, Y_{\rm M}$ at the same
point,  $\bm X \times \bm Y$ is parallel to $\bm e_r$ and, consequently
$\omega_{\rm M}(X_{\rm M}, Y_{\rm M}) \neq 0$.
To show that 
\eqref{OmegaProdDef} is equivalent to \eqref{SimpOmegaDef}, we simply
evaluate components of $\omega$ at  point $\rm M$ determined
by $\bm e_r = \sin \theta(\bm e_x \cos \varphi + \bm e_y \sin \varphi) + \bm e_z \cos \theta$. By remembering that $e_\varphi$ corresponds to
$\sin \theta \bm e_\varphi$, and that $\bm e_\theta \times \bm e_\varphi = \bm e_r$, we get
\begin{align}
    \omega_{ \varphi \theta} \left. \right|_{\rm M} & = \omega(e_\varphi, e_\theta)\left. \right|_{\rm M} = \bm e_r \cdot\Big{(} \sin \theta \bm e_{\varphi} \times \bm e_{\theta}  \Big{)} \nonumber \\
    & = - \sin \theta
\end{align}
in complete agreement with \eqref{SimpOmegaDef}.
Thus, the symplectic form $\omega$ is indeed defined at 
 all points of a sphere and the  apparent vanishing of $\omega$ at poles   is just an artifact of standard spherical coordinates. These problems can be evaded if several coordinate systems are used to cover $\rm S^2$. 

 Finally, we give a few remarks on notation. A coordinate basis of vector fields,
 on a patch of a manifold covered by local coordinates $\{ x^\mu\}$, is commonly denoted by $\{ \partial/\partial x^\mu  \equiv \partial_\mu \}$. The corresponding basis of dual
 vector fields is denoted by $\{  \d x^\mu\}$, so that
 $\d x^\mu(\partial_\nu) = \delta^\mu_\nu$. Using this notation,
 \eqref{HamDif} becomes simply
 $\d H = \partial_1 H \d x^1 + \partial_2 H \d x^2$.
 Also, when working with antisymmetric tensors, one is naturally led to the concept of exterior (or wedge) product --  antisymmetric tensor product defined as
 $\d x^\mu \wedge \d x^\nu:= \d x^\mu \otimes \d x^\nu - \d x^\nu \otimes \d x^\mu$. Using the wedge product, the symplectic form  \eqref{SimpOmegaDef}
 can be written as
 \begin{equation}
     \omega = \sin \theta  \d \theta \wedge \d \varphi.
 \end{equation}
One can easily rewrite all the other tensor fields used in this article
in a similar fashion (See \cite{Frankel, Fecko}).

Most of the phase spaces occurring in classical mechanics are cotangent bundles -- these are 
spaces consisting of an $n$ dimensional manifold (base space) and a copy of $n$ dimensional
dual space $V^*$ attached at each point $\rm M$ of the manifold (typical fiber). The vector space
$V^*$ is the dual space to the space of tangent vectors at each point of the manifold.
An example of cotangent bundle is the simplest phase space, $\mathbb{R}^2$. In this
case, the base manifold is $\mathbb{R}$ and also $V^* = \mathbb{R}$.
Cotangent bundles are $2 n$ dimensional spaces which locally look like $\mathbb{R}^n \times \mathbb{R}^n$ so that each point of cotangent bundle consist 
of a pair $(x({\rm M}), P_{\rm M})$, where the first $n$ entries $x({\rm M}) \in \mathbb{R}^n$ correspond to local
coordinates of the point $\rm M$ and $P_{\rm M} \in \mathbb{R}^n$ is the dual vector $P$ at the point
$\rm M$. The components of dual vector $P$ are interpreted as generalized momenta of 
a given particle configuration and each point of the cotangent  bundle corresponds
to a dynamical state of a particle system. It turns out that cotangent bundles carry natural
symplectic structure \cite{Arnold,Fecko,Frankel}. That is, all cotangent
bundles are true phase spaces. Cotangent bundles also come equipped with a map, called projection, which
assigns to each pair $(x({\rm M}), P_{\rm M})$ the point $\rm M$ where dual vector is located,
$\pi: (x({\rm M}), P_{\rm M}) \mapsto \rm M$. In terms of mechanical
systems, this means that we can always read off a  configuration of  the system (i.e. positions of all the particles) from a given dynamical state.
However, not all phase spaces are cotangent bundles. An important example is
precisely the two sphere. Since $\rm S^2$ is not a cotangent bundle, there is no
projection $\pi$ which would assign a configuration to a given dynamical state.
In this sense the dynamical state of a classical spin cannot be reduced to the dynamics
of a particle, whose position and momentum would correspond to points on $\rm S^2$, and 
classical spin must be treated as a new kind of dynamical system.

By using a different set of coordinates on $\rm S^2$, we may cast
Heisenberg Hamiltonian \eqref{ClassHeisHam} into a form which is more similar
to particle Hamiltonians appearing in classical mechanics.
Indeed, standard spherical coordinates are only one of many possible choices.
Given the fact that $\varphi$ and $\cos \theta$ satisfy canonical Poisson
bracket algebra \eqref{PoissonBracketsSingle}, we may as well use $P=\cos \theta$
as our second phase space coordinate (this choice was already mentioned in Section \ref{ToyS2}). 
Since $\d P = - \d \theta \sin \theta$, the symplectic form $\omega$
reduces to its diagonal form
\begin{equation}
   \omega =  \d \varphi \wedge \d P
\end{equation}
and so does the Poisson tensor \eqref{PoissTenDef}. Since
\begin{equation}
   e_\theta= \frac{\partial}{\partial \theta} = \frac{\partial P}{\partial \theta} \frac{\partial}{\partial P} = - \sin \theta \frac{\partial}{\partial P} = -\sin \theta e_P,
\end{equation}
we have
\begin{align}
    {\mathcal P} & = - \left( \sin \theta \right)^{-1}  \Big{(}  e_{\varphi} \otimes  e_{\theta}
    -  e_{\theta} \otimes  e_{\varphi}\Big{)}\nonumber \\
     & = e_\varphi \otimes e_P - e_P \otimes e_\varphi. \label{PoissTenDefP}
\end{align}
This special set of coordinates which diagonalize $\omega$ are known
as Darboux coordinates \cite{Wood,Fecko}. As it may be anticipated,
the Poisson bracket algebra now takes a particularly simple form
\begin{equation}
     \big{[} \varphi,  P \big{]}_{\rm PB} = {\mathcal P} 
     \Big{(}   \d \varphi, \d P \Big{)} = 1. \label{PBDarb}
\end{equation}
A classical spin vector can also be expressed using Darboux coordinates 
\begin{equation}
    \bm S =  \begin{bmatrix}
       S_x     \\
      S_y         \\
     S_z \\
		\end{bmatrix}
    =
    \begin{bmatrix}
       \sqrt{1-P^2} \cos \varphi     \\
      \sqrt{1-P^2}  \sin \varphi         \\
      P \\
		\end{bmatrix}. \label{ClassSPinComponentsP}
\end{equation}
If we switch to this new coordinate system at each lattice site,
classical Heisenberg Hamiltonian \eqref{ClassHeisHam} becomes
\begin{align}
    H & = -\frac{J}{2} \sum_{\bm n, \bm \lambda} \Bigg[ \sqrt{1-P_{\bm n}^2}  
    \sqrt{1-P_{\bm n+\bm \lambda}^2} \nonumber \\
 &\times \Bigg(\cos(\varphi_{\bm n}) \cos(\varphi_{\bm n + \bm \lambda}) +   
 \sin(\varphi_{\bm n}) \sin(\varphi_{\bm n + \bm \lambda})\Bigg)  \nonumber \\
 & + P_{\bm n} P_{\bm n + \bm \lambda}
     \Bigg]
\end{align}
This Hamiltonian, together with Poisson brackets \eqref{PBDarb}, looks more
like a standard Hamiltonian which describes phase space dynamics of a particle
system -- it contains a term quadratic in momentum $P$, as well as several nonlinear terms
in $\varphi$ and $P$. The downside of this form is that
the $\rm O(3)$ symmetry gets obscured and the Hamiltonian itself looks more complicated.
One should bear in mind that the analogy between classical Heisenberg model and a classical particle systems is an incomplete one. As we already remarked, two sphere $\rm S^2$
is not a cotangent bundle. Therefore, we should not think of $\varphi$ coordinates
as describing positions of particles whose momentum at given position 
is $P$. Rather, the set $\{(\varphi_{\bm n}, P_{\bm n})\}$ specifies a dynamical state of
a classical many-spin system and one cannot interpret  it in terms
of particles' position and momenta.

\section{Summary} \label{SecSumm}

\noindent
Quantum Heisenberg model describes the physics of localized magnetism.
As such, it has many important applications in modeling strongly correlated
electron systems. It  exhibits spontaneous symmetry breaking, emergent 
degrees of freedom carried by quasi-particles, as well as a rich phase diagram
controlled by various coupling constants, anisotropy parameters, etc., which
makes it important from a pedagogical point of view also.
However, it is rarely discussed in the context of quantization 
since Poisson brackets for classical spins  are not of 
a simple form usually taught to undergraduate physics students.
Because of that, Heisenberg model is commonly introduced directly
as a quantum model and classical Heisenberg ferromagnet does not
get a portion of the attention it deserves.

The reason for avoiding detailed discussion on classical Heisenberg model in
introductory statistical mechanics courses lies in the simple fact: The phase space for a single
classical spin is not a Euclidean space and  the standard
definition of Poisson brackets does not apply in this case. We provide here step-by-step
derivation of the Poison bracket algebra of classical spins based on the symplectic 
form on two-sphere.  The presentation is elementary: we  assume only  knowledge
of basic tensor algebra and curvilinear spherical coordinates.
Besides allowing for a discussion on classical spin waves
as collective degrees of freedom which eventually become magnons, arguments presented here put 
classical Heisenberg model into a broader context of quantization
of systems defined on non-Euclidean phase spaces and limitations of standard canonical prescription.

Many excellent books emphasizing the importance of symplectic and Poisson structures
 in contemporary physics  do exist. Besides already mentioned \cite{Fecko,Frankel},
an interested reader could also consult classic texts \cite{Wood,Arnold,MechSymm}
or a recent review \cite{SciPostLN}. Also, Landau-Lifshitz equation
and its solutions are discussed in \cite{MagneticSolitons, Mikeska},
while some references on Monte Carlo simulations of classical Heisenberg model
are \cite{PRBMC,PhysReptsMC,MultiPath,MCJPhysC}.
The physics of Nambu-Goldstone bosons is covered in \cite{BraunerBook}
as well as in \cite{BurgessBook,WeinbergQTF2}.

\begin{acknowledgments}
\noindent
The authors would like to dedicate this paper to students and teachers who
stood against corruption and the collapse of educational system in Serbia
during the academic 2024-2025 year.
We would also like to thank all the taxpayers in Serbia for supporting
science and education. A tiny fraction of  tax revenue is distributed to us by
the Ministry of Science, Technological Development and Innovation of the Republic of Serbia (Grants No. 451-03-137/2025-03/200125 and 451-03-136/2025-03/200125).
    
\end{acknowledgments}

\bibliography{Refs}

\begin{thebibliography}{46}%
\makeatletter
\providecommand \@ifxundefined [1]{%
 \@ifx{#1\undefined}
}%
\providecommand \@ifnum [1]{%
 \ifnum #1\expandafter \@firstoftwo
 \else \expandafter \@secondoftwo
 \fi
}%
\providecommand \@ifx [1]{%
 \ifx #1\expandafter \@firstoftwo
 \else \expandafter \@secondoftwo
 \fi
}%
\providecommand \natexlab [1]{#1}%
\providecommand \enquote  [1]{``#1''}%
\providecommand \bibnamefont  [1]{#1}%
\providecommand \bibfnamefont [1]{#1}%
\providecommand \citenamefont [1]{#1}%
\providecommand \href@noop [0]{\@secondoftwo}%
\providecommand \href [0]{\begingroup \@sanitize@url \@href}%
\providecommand \@href[1]{\@@startlink{#1}\@@href}%
\providecommand \@@href[1]{\endgroup#1\@@endlink}%
\providecommand \@sanitize@url [0]{\catcode `\\12\catcode `\$12\catcode
  `\&12\catcode `\#12\catcode `\^12\catcode `\_12\catcode `\%12\relax}%
\providecommand \@@startlink[1]{}%
\providecommand \@@endlink[0]{}%
\providecommand \url  [0]{\begingroup\@sanitize@url \@url }%
\providecommand \@url [1]{\endgroup\@href {#1}{\urlprefix }}%
\providecommand \urlprefix  [0]{URL }%
\providecommand \Eprint [0]{\href }%
\providecommand \doibase [0]{https://doi.org/}%
\providecommand \selectlanguage [0]{\@gobble}%
\providecommand \bibinfo  [0]{\@secondoftwo}%
\providecommand \bibfield  [0]{\@secondoftwo}%
\providecommand \translation [1]{[#1]}%
\providecommand \BibitemOpen [0]{}%
\providecommand \bibitemStop [0]{}%
\providecommand \bibitemNoStop [0]{.\EOS\space}%
\providecommand \EOS [0]{\spacefactor3000\relax}%
\providecommand \BibitemShut  [1]{\csname bibitem#1\endcsname}%
\let\auto@bib@innerbib\@empty
\bibitem [{\citenamefont {Weinberg}(2012)}]{WeinbergQM}%
  \BibitemOpen
  \bibfield  {author} {\bibinfo {author} {\bibfnamefont {S.}~\bibnamefont
  {Weinberg}},\ }\href
  {https://doi.org/https://doi.org/10.1017/CBO9781316276105} {\emph {\bibinfo
  {title} {Lectures on Quantum Mechanics}}}\ (\bibinfo  {publisher} {Cambridge
  University Press},\ \bibinfo {year} {2012})\BibitemShut {NoStop}%
\bibitem [{\citenamefont {Schiff}(2024)}]{Schiff}%
  \BibitemOpen
  \bibfield  {author} {\bibinfo {author} {\bibfnamefont {L.~I.}\ \bibnamefont
  {Schiff}},\ }\href {https://www.bwpest2018.org/titles/sch68.html} {\emph
  {\bibinfo {title} {Quantum Mechanics}}}\ (\bibinfo  {publisher} {Bow Wow
  Press},\ \bibinfo {year} {2024})\BibitemShut {NoStop}%
\bibitem [{\citenamefont {Ryder}(1996)}]{Ryder}%
  \BibitemOpen
  \bibfield  {author} {\bibinfo {author} {\bibfnamefont {L.}~\bibnamefont
  {Ryder}},\ }\href {https://doi.org/https://doi.org/10.1017/CBO9780511813900}
  {\emph {\bibinfo {title} {Quantum Field Theory}}}\ (\bibinfo  {publisher}
  {Cambridge University Press},\ \bibinfo {year} {1996})\BibitemShut {NoStop}%
\bibitem [{\citenamefont {Weinberg}(2008)}]{WeinbergQTF1}%
  \BibitemOpen
  \bibfield  {author} {\bibinfo {author} {\bibfnamefont {S.}~\bibnamefont
  {Weinberg}},\ }\href
  {http://www.cambridge.org/us/academic/subjects/physics/theoretical-physics-and-mathematical-physics/quantum-theory-fields-volume-1}
  {\emph {\bibinfo {title} {The Quantum Theory of Fields, Vol. I}}}\ (\bibinfo
  {publisher} {Cambridge University Press},\ \bibinfo {year}
  {2008})\BibitemShut {NoStop}%
\bibitem [{\citenamefont {Auerbach}(1994)}]{Auerbach}%
  \BibitemOpen
  \bibfield  {author} {\bibinfo {author} {\bibfnamefont {A.}~\bibnamefont
  {Auerbach}},\ }\href
  {https://doi.org/https://doi.org/10.1007/978-1-4612-0869-3} {\emph {\bibinfo
  {title} {Interacting Electrons and Quantum Magnetism}}}\ (\bibinfo
  {publisher} {Springer},\ \bibinfo {year} {1994})\BibitemShut {NoStop}%
\bibitem [{\citenamefont {Kittel}\ and\ \citenamefont {McEuen}(2018)}]{Kittel}%
  \BibitemOpen
  \bibfield  {author} {\bibinfo {author} {\bibfnamefont {C.}~\bibnamefont
  {Kittel}}\ and\ \bibinfo {author} {\bibfnamefont {P.}~\bibnamefont
  {McEuen}},\ }\href
  {https://www.wiley.com/en-us/Introduction+to+Solid+State+Physics%2C+8th+Edition-p-9780471415268}
  {\emph {\bibinfo {title} {Introduction to Solid State Physics}}}\ (\bibinfo
  {publisher} {John Wiley \& Sons},\ \bibinfo {year} {2018})\BibitemShut
  {NoStop}%
\bibitem [{\citenamefont {Nolting}\ and\ \citenamefont
  {Ramakanth}(2009)}]{Nolting}%
  \BibitemOpen
  \bibfield  {author} {\bibinfo {author} {\bibfnamefont {W.}~\bibnamefont
  {Nolting}}\ and\ \bibinfo {author} {\bibfnamefont {A.}~\bibnamefont
  {Ramakanth}},\ }\href
  {https://doi.org/https://doi.org/10.1007/978-3-540-85416-6} {\emph {\bibinfo
  {title} {Quantum Theory of Magnetism}}}\ (\bibinfo  {publisher} {Springer},\
  \bibinfo {year} {2009})\BibitemShut {NoStop}%
\bibitem [{\citenamefont {Altland}\ and\ \citenamefont {Simons}(2023)}]{CMFT}%
  \BibitemOpen
  \bibfield  {author} {\bibinfo {author} {\bibfnamefont {A.}~\bibnamefont
  {Altland}}\ and\ \bibinfo {author} {\bibfnamefont {B.}~\bibnamefont
  {Simons}},\ }\href {https://doi.org/https://doi.org/10.1017/9781108781244}
  {\emph {\bibinfo {title} {Condensed Matter Field Theory}}}\ (\bibinfo
  {publisher} {Cambridge University Press},\ \bibinfo {year}
  {2023})\BibitemShut {NoStop}%
\bibitem [{\citenamefont {Mattis}(1981)}]{Mattis}%
  \BibitemOpen
  \bibfield  {author} {\bibinfo {author} {\bibfnamefont {D.}~\bibnamefont
  {Mattis}},\ }\href {https://link.springer.com/book/10.1007/978-3-642-83238-3}
  {\emph {\bibinfo {title} {The Theory of Magnetism I -- Statics and
  Dynamics}}}\ (\bibinfo  {publisher} {Springer},\ \bibinfo {year}
  {1981})\BibitemShut {NoStop}%
\bibitem [{\citenamefont {Yosida}(1996)}]{Yosida}%
  \BibitemOpen
  \bibfield  {author} {\bibinfo {author} {\bibfnamefont {K.}~\bibnamefont
  {Yosida}},\ }\href {https://link.springer.com/book/9783540606512} {\emph
  {\bibinfo {title} {Theory of Magnetism}}}\ (\bibinfo  {publisher}
  {Springer},\ \bibinfo {year} {1996})\BibitemShut {NoStop}%
\bibitem [{\citenamefont {Peskin}\ and\ \citenamefont
  {Schroeder}(2018)}]{Peskin}%
  \BibitemOpen
  \bibfield  {author} {\bibinfo {author} {\bibfnamefont {M.~E.}\ \bibnamefont
  {Peskin}}\ and\ \bibinfo {author} {\bibfnamefont {D.~V.}\ \bibnamefont
  {Schroeder}},\ }\href {https://doi.org/https://doi.org/10.1201/9780429503559}
  {\emph {\bibinfo {title} {An Introduction to Quantum Field Theory}}}\
  (\bibinfo  {publisher} {CRC press},\ \bibinfo {year} {2018})\BibitemShut
  {NoStop}%
\bibitem [{\citenamefont {Ryder}(2012)}]{RyderGR}%
  \BibitemOpen
  \bibfield  {author} {\bibinfo {author} {\bibfnamefont {L.}~\bibnamefont
  {Ryder}},\ }\href {https://doi.org/https://doi.org/10.1017/CBO9780511809033}
  {\emph {\bibinfo {title} {Introduction to General Relativity}}}\ (\bibinfo
  {publisher} {Cambridge University Press},\ \bibinfo {year}
  {2012})\BibitemShut {NoStop}%
\bibitem [{\citenamefont {Zee}(2013)}]{ZeeGR}%
  \BibitemOpen
  \bibfield  {author} {\bibinfo {author} {\bibfnamefont {A.}~\bibnamefont
  {Zee}},\ }\href
  {https://press.princeton.edu/books/hardcover/9780691145587/einstein-gravity-in-a-nutshell?srsltid=AfmBOopMOUm_B3AZw3jK7CYWrV7yL0SC6-d52ybVecU8lZmfkynbKhns}
  {\emph {\bibinfo {title} {Einstein Gravity in a Nutshell}}}\ (\bibinfo
  {publisher} {Princeton University Press},\ \bibinfo {year}
  {2013})\BibitemShut {NoStop}%
\bibitem [{\citenamefont {Fecko}(2006)}]{Fecko}%
  \BibitemOpen
  \bibfield  {author} {\bibinfo {author} {\bibfnamefont {M.}~\bibnamefont
  {Fecko}},\ }\href {https://doi.org/https://doi.org/10.1017/CBO9780511755590}
  {\emph {\bibinfo {title} {Differential Geometry and Lie Groups for
  Physicists}}}\ (\bibinfo  {publisher} {Cambridge University Press},\ \bibinfo
  {year} {2006})\BibitemShut {NoStop}%
\bibitem [{\citenamefont {Frankel}(2012)}]{Frankel}%
  \BibitemOpen
  \bibfield  {author} {\bibinfo {author} {\bibfnamefont {T.}~\bibnamefont
  {Frankel}},\ }\href
  {https://doi.org/https://doi.org/10.1017/CBO9781139061377} {\emph {\bibinfo
  {title} {The Geometry of Physics}}}\ (\bibinfo  {publisher} {Cambridge
  University Press},\ \bibinfo {year} {2012})\BibitemShut {NoStop}%
\bibitem [{\citenamefont {Halmos}(2020)}]{Halmos}%
  \BibitemOpen
  \bibfield  {author} {\bibinfo {author} {\bibfnamefont {P.}~\bibnamefont
  {Halmos}},\ }\href {https://www.bwpest2018.org/titles/hal58.html} {\emph
  {\bibinfo {title} {Linear Algebra: Finite-Dimensional Vector Spaces}}}\
  (\bibinfo  {publisher} {Bow Wow Press},\ \bibinfo {year} {2020})\BibitemShut
  {NoStop}%
\bibitem [{\citenamefont {Axler}(2024)}]{Axler}%
  \BibitemOpen
  \bibfield  {author} {\bibinfo {author} {\bibfnamefont {S.}~\bibnamefont
  {Axler}},\ }\href {https://doi.org/https://doi.org/10.1007/978-3-031-41026-0}
  {\emph {\bibinfo {title} {Linear Algebra Done Right}}}\ (\bibinfo
  {publisher} {Springer},\ \bibinfo {year} {2024})\BibitemShut {NoStop}%
\bibitem [{\citenamefont {Hassani}(2013)}]{Hasani}%
  \BibitemOpen
  \bibfield  {author} {\bibinfo {author} {\bibfnamefont {S.}~\bibnamefont
  {Hassani}},\ }\href {https://doi.org/10.1007/978-3-319-01195-0} {\emph
  {\bibinfo {title} {Mathematical Physics}}}\ (\bibinfo  {publisher}
  {{Springer}},\ \bibinfo {year} {2013})\BibitemShut {NoStop}%
\bibitem [{\citenamefont {Arfken}\ \emph {et~al.}(2013)\citenamefont {Arfken},
  \citenamefont {Weber},\ and\ \citenamefont {Harris}}]{AWH}%
  \BibitemOpen
  \bibfield  {author} {\bibinfo {author} {\bibfnamefont {G.}~\bibnamefont
  {Arfken}}, \bibinfo {author} {\bibfnamefont {H.}~\bibnamefont {Weber}},\ and\
  \bibinfo {author} {\bibfnamefont {F.~E.}\ \bibnamefont {Harris}},\ }\href
  {https://doi.org/https://doi.org/10.1016/C2009-0-30629-7} {\emph {\bibinfo
  {title} {Mathematical Methods for Physicists}}}\ (\bibinfo  {publisher}
  {Academic Press},\ \bibinfo {year} {2013})\BibitemShut {NoStop}%
\bibitem [{\citenamefont {Witten}(1984)}]{Witten}%
  \BibitemOpen
  \bibfield  {author} {\bibinfo {author} {\bibfnamefont {E.}~\bibnamefont
  {Witten}},\ }\bibfield  {title} {\bibinfo {title} {Non-abelian bosonization
  in two dimensions},\ }\href
  {https://doi.org/https://doi.org/10.1007/BF01215276} {\bibfield  {journal}
  {\bibinfo  {journal} {Communications in Mathematical Physics}\ }\textbf
  {\bibinfo {volume} {92}},\ \bibinfo {pages} {455} (\bibinfo {year}
  {1984})}\BibitemShut {NoStop}%
\bibitem [{\citenamefont {Marsden}\ and\ \citenamefont
  {Ratiu}(1999)}]{MechSymm}%
  \BibitemOpen
  \bibfield  {author} {\bibinfo {author} {\bibfnamefont {J.}~\bibnamefont
  {Marsden}}\ and\ \bibinfo {author} {\bibfnamefont {T.}~\bibnamefont
  {Ratiu}},\ }\href {https://doi.org/https://doi.org/10.1007/978-0-387-21792-5}
  {\emph {\bibinfo {title} {{Introduction to Mechanics and Symmetry}}}}\
  (\bibinfo  {publisher} {Springer},\ \bibinfo {year} {1999})\BibitemShut
  {NoStop}%
\bibitem [{\citenamefont {Aldaya}\ \emph {et~al.}(2009)\citenamefont {Aldaya},
  \citenamefont {Calixto}, \citenamefont {Guerrero},\ and\ \citenamefont
  {López-Ruiz}}]{GuerreroS2}%
  \BibitemOpen
  \bibfield  {author} {\bibinfo {author} {\bibfnamefont {V.}~\bibnamefont
  {Aldaya}}, \bibinfo {author} {\bibfnamefont {M.}~\bibnamefont {Calixto}},
  \bibinfo {author} {\bibfnamefont {J.}~\bibnamefont {Guerrero}},\ and\
  \bibinfo {author} {\bibfnamefont {F.}~\bibnamefont {López-Ruiz}},\
  }\bibfield  {title} {\bibinfo {title} {{Group-quantization of nonlinear sigma
  models: Particle on $\rm S^2$ revisited}},\ }\href
  {https://doi.org/https://doi.org/10.1016/S0034-4877(09)90019-2} {\bibfield
  {journal} {\bibinfo  {journal} {Reports on Mathematical Physics}\ }\textbf
  {\bibinfo {volume} {64}},\ \bibinfo {pages} {49} (\bibinfo {year}
  {2009})}\BibitemShut {NoStop}%
\bibitem [{\citenamefont {e~Silva}\ and\ \citenamefont
  {Jacobson}(2021)}]{SilvaJPA}%
  \BibitemOpen
  \bibfield  {author} {\bibinfo {author} {\bibfnamefont {R.~A.}\ \bibnamefont
  {e~Silva}}\ and\ \bibinfo {author} {\bibfnamefont {T.}~\bibnamefont
  {Jacobson}},\ }\bibfield  {title} {\bibinfo {title} {Particle on the sphere:
  group-theoretic quantization in the presence of a magnetic monopole},\ }\href
  {https://doi.org/10.1088/1751-8121/abf961} {\bibfield  {journal} {\bibinfo
  {journal} {Journal of Physics A: Mathematical and Theoretical}\ }\textbf
  {\bibinfo {volume} {54}},\ \bibinfo {pages} {235303} (\bibinfo {year}
  {2021})}\BibitemShut {NoStop}%
\bibitem [{\citenamefont {Stone}(1989)}]{Stone}%
  \BibitemOpen
  \bibfield  {author} {\bibinfo {author} {\bibfnamefont {M.}~\bibnamefont
  {Stone}},\ }\bibfield  {title} {\bibinfo {title} {Supersymmetry and the
  quantum mechanics of spin},\ }\href
  {https://doi.org/https://doi.org/10.1016/0550-3213(89)90408-2} {\bibfield
  {journal} {\bibinfo  {journal} {Nuclear Physics B}\ }\textbf {\bibinfo
  {volume} {314}},\ \bibinfo {pages} {557} (\bibinfo {year}
  {1989})}\BibitemShut {NoStop}%
\bibitem [{\citenamefont {Burgess}(2000)}]{Burgess}%
  \BibitemOpen
  \bibfield  {author} {\bibinfo {author} {\bibfnamefont {C.}~\bibnamefont
  {Burgess}},\ }\bibfield  {title} {\bibinfo {title} {Goldstone and
  pseudo-goldstone bosons in nuclear, particle and condensed-matter physics},\
  }\href {https://doi.org/10.1016/S0370-1573(99)00111-8} {\bibfield  {journal}
  {\bibinfo  {journal} {Phys. Repts.}\ }\textbf {\bibinfo {volume} {330}},\
  \bibinfo {pages} {193 } (\bibinfo {year} {2000})}\BibitemShut {NoStop}%
\bibitem [{\citenamefont {Holm}\ and\ \citenamefont {Janke}(1993)}]{Holm}%
  \BibitemOpen
  \bibfield  {author} {\bibinfo {author} {\bibfnamefont {C.}~\bibnamefont
  {Holm}}\ and\ \bibinfo {author} {\bibfnamefont {W.}~\bibnamefont {Janke}},\
  }\bibfield  {title} {\bibinfo {title} {Critical exponents of the classical
  three-dimensional heisenberg model: A single-cluster monte carlo study},\
  }\href {https://doi.org/10.1103/PhysRevB.48.936} {\bibfield  {journal}
  {\bibinfo  {journal} {Phys. Rev. B}\ }\textbf {\bibinfo {volume}
  {\textbf{48}}},\ \bibinfo {pages} {936} (\bibinfo {year} {1993})}\BibitemShut
  {NoStop}%
\bibitem [{\citenamefont {Crnkovic}\ and\ \citenamefont
  {Witten}(1987)}]{CrnkovicWitten}%
  \BibitemOpen
  \bibfield  {author} {\bibinfo {author} {\bibfnamefont {C.}~\bibnamefont
  {Crnkovic}}\ and\ \bibinfo {author} {\bibfnamefont {E.}~\bibnamefont
  {Witten}},\ }\bibfield  {title} {\bibinfo {title} {Covariant description of
  canonical formalism in geometrical theories},\ }in\ \href
  {https://www.cambridge.org/us/universitypress/subjects/physics/cosmology-relativity-and-gravitation/three-hundred-years-gravitation?format=PB}
  {\emph {\bibinfo {booktitle} {Three hundred years of gravitation}}},\
  \bibinfo {editor} {edited by\ \bibinfo {editor} {\bibfnamefont {S.~W.}\
  \bibnamefont {Hawking}}\ and\ \bibinfo {editor} {\bibfnamefont
  {W.}~\bibnamefont {Israel}}}\ (\bibinfo {year} {1987})\ pp.\ \bibinfo {pages}
  {676--684}\BibitemShut {NoStop}%
\bibitem [{\citenamefont {Villain}(1974)}]{Villain}%
  \BibitemOpen
  \bibfield  {author} {\bibinfo {author} {\bibfnamefont {J.}~\bibnamefont
  {Villain}},\ }\bibfield  {title} {\bibinfo {title} {Quantum theory of one-and
  two-dimensional ferro-and antiferromagnets with an easy magnetization plane.
  i. ideal 1-d or 2-d lattices without in-plane anisotropy},\ }\href
  {https://doi.org/https://doi.org/10.1051/jphys:0197400350102700} {\bibfield
  {journal} {\bibinfo  {journal} {Journal de Physique}\ }\textbf {\bibinfo
  {volume} {35}},\ \bibinfo {pages} {27} (\bibinfo {year} {1974})}\BibitemShut
  {NoStop}%
\bibitem [{\citenamefont {Bulgac}\ and\ \citenamefont
  {Kusnezov}(1990)}]{ClasLimit}%
  \BibitemOpen
  \bibfield  {author} {\bibinfo {author} {\bibfnamefont {A.}~\bibnamefont
  {Bulgac}}\ and\ \bibinfo {author} {\bibfnamefont {D.}~\bibnamefont
  {Kusnezov}},\ }\bibfield  {title} {\bibinfo {title} {Classical limit for lie
  algebras},\ }\href
  {https://doi.org/https://doi.org/10.1016/0003-4916(90)90373-V} {\bibfield
  {journal} {\bibinfo  {journal} {Annals of Physics}\ }\textbf {\bibinfo
  {volume} {199}},\ \bibinfo {pages} {187} (\bibinfo {year}
  {1990})}\BibitemShut {NoStop}%
\bibitem [{\citenamefont {Brauner}(2024)}]{BraunerBook}%
  \BibitemOpen
  \bibfield  {author} {\bibinfo {author} {\bibfnamefont {T.}~\bibnamefont
  {Brauner}},\ }\href {https://doi.org/10.1007/978-3-031-48378-3} {\emph
  {\bibinfo {title} {{Effective Field Theory for Spontaneously Broken
  Symmetry}}}}\ (\bibinfo  {publisher} {Springer},\ \bibinfo {year}
  {2024})\BibitemShut {NoStop}%
\bibitem [{\citenamefont {Yang}\ and\ \citenamefont
  {Hirschfelder}(1980)}]{YangPB}%
  \BibitemOpen
  \bibfield  {author} {\bibinfo {author} {\bibfnamefont {K.-H.}\ \bibnamefont
  {Yang}}\ and\ \bibinfo {author} {\bibfnamefont {J.~O.}\ \bibnamefont
  {Hirschfelder}},\ }\bibfield  {title} {\bibinfo {title} {Generalizations of
  classical poisson brackets to include spin},\ }\href
  {https://doi.org/10.1103/PhysRevA.22.1814} {\bibfield  {journal} {\bibinfo
  {journal} {Phys. Rev. A}\ }\textbf {\bibinfo {volume} {22}},\ \bibinfo
  {pages} {1814} (\bibinfo {year} {1980})}\BibitemShut {NoStop}%
\bibitem [{\citenamefont {Rado\v{s}evi\'{c}}(2015)}]{AnnPhys2015}%
  \BibitemOpen
  \bibfield  {author} {\bibinfo {author} {\bibfnamefont {S.~M.}\ \bibnamefont
  {Rado\v{s}evi\'{c}}},\ }\bibfield  {title} {\bibinfo {title}
  {{Magnon–magnon interactions in O(3) ferromagnets and equations of motion
  for spin operators}},\ }\href
  {https://doi.org/https://doi.org/10.1016/j.aop.2015.08.003} {\bibfield
  {journal} {\bibinfo  {journal} {Annals of Physics}\ }\textbf {\bibinfo
  {volume} {362}},\ \bibinfo {pages} {336} (\bibinfo {year}
  {2015})}\BibitemShut {NoStop}%
\bibitem [{\citenamefont {Lakshmanan}(2011)}]{LLreview}%
  \BibitemOpen
  \bibfield  {author} {\bibinfo {author} {\bibfnamefont {M.}~\bibnamefont
  {Lakshmanan}},\ }\bibfield  {title} {\bibinfo {title} {The fascinating world
  of the landau–lifshitz–gilbert equation: an overview},\ }\href
  {https://doi.org/10.1098/rsta.2010.0319} {\bibfield  {journal} {\bibinfo
  {journal} {Philosophical Transactions of the Royal Society A: Mathematical,
  Physical and Engineering Sciences}\ }\textbf {\bibinfo {volume} {369}},\
  \bibinfo {pages} {1280} (\bibinfo {year} {2011})}\BibitemShut {NoStop}%
\bibitem [{\citenamefont {Watanabe}\ and\ \citenamefont
  {Murayama}(2014)}]{PRX}%
  \BibitemOpen
  \bibfield  {author} {\bibinfo {author} {\bibfnamefont {H.}~\bibnamefont
  {Watanabe}}\ and\ \bibinfo {author} {\bibfnamefont {H.}~\bibnamefont
  {Murayama}},\ }\bibfield  {title} {\bibinfo {title} {{Effective Lagrangian
  for nonrelativistic systems}},\ }\href
  {https://doi.org/10.1103/PhysRevX.4.031057} {\bibfield  {journal} {\bibinfo
  {journal} {Phys. Rev. X}\ }\textbf {\bibinfo {volume} {\textbf{4}}},\
  \bibinfo {pages} {031057} (\bibinfo {year} {2014})}\BibitemShut {NoStop}%
\bibitem [{\citenamefont {Rado{\v{s}}evi{\'c}}(2025)}]{ArXiv2024}%
  \BibitemOpen
  \bibfield  {author} {\bibinfo {author} {\bibfnamefont {S.}~\bibnamefont
  {Rado{\v{s}}evi{\'c}}},\ }\bibfield  {title} {\bibinfo {title} {{Geometry of
  classical Nambu-Goldstone fields}},\ }\href
  {https://doi.org/https://doi.org/10.1016/j.aop.2025.169931} {\bibfield
  {journal} {\bibinfo  {journal} {Annals of Physics}\ }\textbf {\bibinfo
  {volume} {474}},\ \bibinfo {pages} {169931} (\bibinfo {year}
  {2025})}\BibitemShut {NoStop}%
\bibitem [{\citenamefont {Arnol'd}(2013)}]{Arnold}%
  \BibitemOpen
  \bibfield  {author} {\bibinfo {author} {\bibfnamefont {V.~I.}\ \bibnamefont
  {Arnol'd}},\ }\href
  {https://doi.org/https://doi.org/10.1007/978-1-4757-2063-1} {\emph {\bibinfo
  {title} {Mathematical Methods of Classical Mechanics}}}\ (\bibinfo
  {publisher} {Springer},\ \bibinfo {year} {2013})\BibitemShut {NoStop}%
\bibitem [{\citenamefont {Woodhouse}(1992)}]{Wood}%
  \BibitemOpen
  \bibfield  {author} {\bibinfo {author} {\bibfnamefont {N.~M.~J.}\
  \bibnamefont {Woodhouse}},\ }\href@noop {} {\emph {\bibinfo {title}
  {\href{https://global.oup.com/academic/product/geometric-quantization-9780198502708?cc=rs&lang=en&}{Geometric
  quantization}}}}\ (\bibinfo  {publisher} {Oxford university press},\ \bibinfo
  {year} {1992})\BibitemShut {NoStop}%
\bibitem [{\citenamefont {Gieres}(2023)}]{SciPostLN}%
  \BibitemOpen
  \bibfield  {author} {\bibinfo {author} {\bibfnamefont {F.}~\bibnamefont
  {Gieres}},\ }\bibfield  {title} {\bibinfo {title} {{Covariant canonical
  formulations of classical field theories}},\ }\href
  {https://doi.org/10.21468/SciPostPhysLectNotes.77} {\bibfield  {journal}
  {\bibinfo  {journal} {SciPost Phys. Lect. Notes}\ ,\ \bibinfo {pages}
  {\textbf{77}}} (\bibinfo {year} {2023})}\BibitemShut {NoStop}%
\bibitem [{\citenamefont {Kosevich}\ \emph {et~al.}(1990)\citenamefont
  {Kosevich}, \citenamefont {Ivanov},\ and\ \citenamefont
  {Kovalev}}]{MagneticSolitons}%
  \BibitemOpen
  \bibfield  {author} {\bibinfo {author} {\bibfnamefont {A.}~\bibnamefont
  {Kosevich}}, \bibinfo {author} {\bibfnamefont {B.}~\bibnamefont {Ivanov}},\
  and\ \bibinfo {author} {\bibfnamefont {A.}~\bibnamefont {Kovalev}},\
  }\bibfield  {title} {\bibinfo {title} {Magnetic solitons},\ }\href
  {https://doi.org/https://doi.org/10.1016/0370-1573(90)90130-T} {\bibfield
  {journal} {\bibinfo  {journal} {Physics Reports}\ }\textbf {\bibinfo {volume}
  {194}},\ \bibinfo {pages} {117} (\bibinfo {year} {1990})}\BibitemShut
  {NoStop}%
\bibitem [{\citenamefont {Mikeska}\ and\ \citenamefont
  {Steiner}(1991)}]{Mikeska}%
  \BibitemOpen
  \bibfield  {author} {\bibinfo {author} {\bibfnamefont {H.-J.}\ \bibnamefont
  {Mikeska}}\ and\ \bibinfo {author} {\bibfnamefont {M.}~\bibnamefont
  {Steiner}},\ }\bibfield  {title} {\bibinfo {title} {Solitary excitations in
  one-dimensional magnets},\ }\href {https://doi.org/10.1080/00018739100101492}
  {\bibfield  {journal} {\bibinfo  {journal} {Advances in Physics}\ }\textbf
  {\bibinfo {volume} {40}},\ \bibinfo {pages} {191} (\bibinfo {year}
  {1991})}\BibitemShut {NoStop}%
\bibitem [{\citenamefont {Chen}\ \emph {et~al.}(1993)\citenamefont {Chen},
  \citenamefont {Ferrenberg},\ and\ \citenamefont {Landau}}]{PRBMC}%
  \BibitemOpen
  \bibfield  {author} {\bibinfo {author} {\bibfnamefont {K.}~\bibnamefont
  {Chen}}, \bibinfo {author} {\bibfnamefont {A.~M.}\ \bibnamefont
  {Ferrenberg}},\ and\ \bibinfo {author} {\bibfnamefont {D.~P.}\ \bibnamefont
  {Landau}},\ }\bibfield  {title} {\bibinfo {title} {Static critical behavior
  of three-dimensional classical heisenberg models: A high-resolution monte
  carlo study},\ }\href {https://doi.org/10.1103/PhysRevB.48.3249} {\bibfield
  {journal} {\bibinfo  {journal} {Phys. Rev. B}\ }\textbf {\bibinfo {volume}
  {48}},\ \bibinfo {pages} {3249} (\bibinfo {year} {1993})}\BibitemShut
  {NoStop}%
\bibitem [{\citenamefont {Pelissetto}\ and\ \citenamefont
  {Vicari}(2002)}]{PhysReptsMC}%
  \BibitemOpen
  \bibfield  {author} {\bibinfo {author} {\bibfnamefont {A.}~\bibnamefont
  {Pelissetto}}\ and\ \bibinfo {author} {\bibfnamefont {E.}~\bibnamefont
  {Vicari}},\ }\bibfield  {title} {\bibinfo {title} {Critical phenomena and
  renormalization-group theory},\ }\href
  {https://doi.org/https://doi.org/10.1016/S0370-1573(02)00219-3} {\bibfield
  {journal} {\bibinfo  {journal} {Physics Reports}\ }\textbf {\bibinfo {volume}
  {368}},\ \bibinfo {pages} {549} (\bibinfo {year} {2002})}\BibitemShut
  {NoStop}%
\bibitem [{\citenamefont {Rakić}\ \emph {et~al.}(2016)\citenamefont {Rakić},
  \citenamefont {Radošević}, \citenamefont {Mali}, \citenamefont
  {Stričević},\ and\ \citenamefont {Petrić}}]{MultiPath}%
  \BibitemOpen
  \bibfield  {author} {\bibinfo {author} {\bibfnamefont {P.~S.}\ \bibnamefont
  {Rakić}}, \bibinfo {author} {\bibfnamefont {S.~M.}\ \bibnamefont
  {Radošević}}, \bibinfo {author} {\bibfnamefont {P.~M.}\ \bibnamefont
  {Mali}}, \bibinfo {author} {\bibfnamefont {L.~M.}\ \bibnamefont
  {Stričević}},\ and\ \bibinfo {author} {\bibfnamefont {T.~D.}\ \bibnamefont
  {Petrić}},\ }\bibfield  {title} {\bibinfo {title} {Multipath metropolis
  simulation: An application to the classical heisenberg model},\ }\href
  {https://doi.org/https://doi.org/10.1016/j.physa.2015.08.038} {\bibfield
  {journal} {\bibinfo  {journal} {Physica A: Statistical Mechanics and its
  Applications}\ }\textbf {\bibinfo {volume} {441}},\ \bibinfo {pages} {69}
  (\bibinfo {year} {2016})}\BibitemShut {NoStop}%
\bibitem [{\citenamefont {Alzate-Cardona}\ \emph {et~al.}(2019)\citenamefont
  {Alzate-Cardona}, \citenamefont {Sabogal-Suárez}, \citenamefont {Evans},\
  and\ \citenamefont {Restrepo-Parra}}]{MCJPhysC}%
  \BibitemOpen
  \bibfield  {author} {\bibinfo {author} {\bibfnamefont {J.~D.}\ \bibnamefont
  {Alzate-Cardona}}, \bibinfo {author} {\bibfnamefont {D.}~\bibnamefont
  {Sabogal-Suárez}}, \bibinfo {author} {\bibfnamefont {R.~F.~L.}\ \bibnamefont
  {Evans}},\ and\ \bibinfo {author} {\bibfnamefont {E.}~\bibnamefont
  {Restrepo-Parra}},\ }\bibfield  {title} {\bibinfo {title} {Optimal phase
  space sampling for monte carlo simulations of heisenberg spin systems},\
  }\href {https://doi.org/10.1088/1361-648X/aaf852} {\bibfield  {journal}
  {\bibinfo  {journal} {Journal of Physics: Condensed Matter}\ }\textbf
  {\bibinfo {volume} {31}},\ \bibinfo {pages} {095802} (\bibinfo {year}
  {2019})}\BibitemShut {NoStop}%
\bibitem [{\citenamefont {Burgess}(2020)}]{BurgessBook}%
  \BibitemOpen
  \bibfield  {author} {\bibinfo {author} {\bibfnamefont {C.~P.}\ \bibnamefont
  {Burgess}},\ }\href {https://doi.org/https://doi.org/10.1017/9781139048040}
  {\emph {\bibinfo {title} {Introduction to effective field theory}}}\
  (\bibinfo  {publisher} {Cambridge University Press},\ \bibinfo {year}
  {2020})\BibitemShut {NoStop}%
\bibitem [{\citenamefont {Weinberg}(2010)}]{WeinbergQTF2}%
  \BibitemOpen
  \bibfield  {author} {\bibinfo {author} {\bibfnamefont {S.}~\bibnamefont
  {Weinberg}},\ }\href
  {http://www.cambridge.org/us/academic/subjects/physics/theoretical-physics-and-mathematical-physics/quantum-theory-fields-volume-2}
  {\emph {\bibinfo {title} {The Quantum Theory of Fields, Vol. II}}}\ (\bibinfo
   {publisher} {Cambridge University Press},\ \bibinfo {year}
  {2010})\BibitemShut {NoStop}%
\end{thebibliography}%

\end{document}